\documentclass{elsarticle}

\usepackage[utf8]{inputenc}

\usepackage{tabularx}
\usepackage{booktabs}
\usepackage{multirow}

\usepackage{amsmath}
\usepackage{hyperref}

\usepackage{subfig}

\usepackage{paralist}

\usepackage{cleveref}

\def\copyrightspace{
\long\def\@makefntext##1{\noindent ##1}
\footnotesep 1em
\footnotetext[0]{\em Copyright \copyright\ 2020 for this paper by its authors. Use permitted under Creative Commons License Attribution 4.0 International (CC BY 4.0).}
}

\makeatletter
\def\algbackskip{\hskip-\ALG@thistlm}

\def\ps@pprintTitle{%
 \let\@oddhead\@empty
 \let\@evenhead\@empty
 \def\@oddfoot{ 
 }%
 \let\@evenfoot\@oddfoot}

\makeatother

\begin{document}
\title{Views on Quality Requirements in Academia and Practice: Commonalities, Differences, and Context-Dependent Grey Areas}

\author[tub]{Andreas Vogelsang}
\ead{andreas.vogelsang@tu-berlin.de}
\author[tab]{Jonas Eckhardt}
\ead{jonaseckhardt@googlemail.com}
\author[bth]{Daniel Mendez}
\ead{daniel.mendez@bth.se}
\author[ubo]{Moritz Berger}
\ead{moritz.berger@imbie.uni-bonn.de}

\address[tub]{Technische Universit{\"a}t Berlin, Germany}
\address[tab]{Tableau Software, Germany}
\address[bth]{Blekinge Institute of Technology, Sweden, and fortiss GmbH, Germany}
\address[ubo]{Universit{\"a}t Bonn, Germany}

\fntext[fn1]{This is the Accepted Manuscript of: \\
Vogelsang, A., Eckhardt, J, Mendez, D., Berger, M. (2020): Views on Quality Requirements in Academia and Practice: Commonalities, Differences, and Context-Dependent Grey Areas. Information and Software Technology. \url{https://doi.org/10.1016/j.infsof.2019.106253}\\
This work is licensed under a Creative Commons Attribution-NonCommercial-NoDerivatives 4.0 International License, \url{http://creativecommons.org/licenses/by-nc-nd/4.0/}.}

\begin{abstract}
\textbf{Context:}
Quality requirements (QRs) are a topic of constant discussions both in industry and academia. Debates entwine around the definition of quality requirements, the way how to handle them, or their importance for project success. 
While many academic endeavors contribute to the body of knowledge about QRs, 
practitioners may have different views. In fact, we still lack a consistent body of knowledge on QRs since much of the discussion around this topic is still dominated by 
observations that are strongly context-dependent. This holds for both academic and practitioners' views. Our assumption is that, in consequence, those views may differ.

\noindent\textbf{Objective:}
We report on a study to better understand the extent to which available research statements on quality requirements, as found in exemplary peer-reviewed and frequently cited publications, are reflected in the perception of practitioners. Our goal is to analyze differences, commonalities, and context-dependent grey areas in the views of academics and practitioners to allow a discussion on potential misconceptions (on either sides) and opportunities for future research.

\noindent\textbf{Method:}
We conducted a survey with 109 practitioners to assess whether they agree with research statements about QRs reflected in the literature. Based on a statistical model, we evaluate the impact of a set of context factors to the perception of research statements. 

\noindent\textbf{Results:}
Our results show that a majority of the statements is well respected by practitioners; however, not all of them. When examining the different groups and backgrounds of respondents, we noticed interesting deviations of perceptions within different groups that may lead to new research questions.

\noindent\textbf{Conclusions:}
Our results help identifying prevalent context-dependent differences about how academics and practitioners view QRs and pinpointing statements where further research might be useful.

\end{abstract}


\maketitle

\section{Introduction}
Requirements Engineering (RE) constitutes an important success factor in today's software development projects and is often understood as a determinant of productivity and (product) quality~\cite{DC06}. Despite its importance, the discipline remains difficult to investigate due to its inherent complexity and its dependency to the various influences by the particularities of industrial sectors, domains, and individual project environments. This holds especially for quality requirements (QR), what they are, and how they should be handled in practice. Quality requirements are often considered separately to functional requirements in research~\cite{robertson2012mastering,Sommerville97} and practice~\cite{Ameller12,borg2003bad,chung1995dealing,svensson2009quality} alike as they tend to be treated differently along their elicitation, documentation, and validation. Yet, there is still no common agreement about what quality requirements exactly are~\cite{Glinz07} despite the fact that there is even an ISO standard~\cite{iso25010} with a definition. The discourse about quality requirements is still dominated even by the question how to differentiate them from functional requirements~\cite{Broy16Rethinking, Glinz07}. 

In empirically-informed work~\cite{MW+16,Mendez-Passoth18}, we pointed out that conceptual contributions to RE are still heavily steered by conventional wisdom rather than by empirical observations. In fact, we still do not exploit the full potential of empirical software engineering principles in RE to reveal robust theories in tune with practical problems. It is per se difficult to provide proper empirical figures that could demonstrate, for instance, how theoretical concepts are reflected in industrial practices or, the other way around, how industry practitioners' experiences, observations, and opinions can find their way back into academic contributions. This existing gap between academia and industry holds especially for fuzzy notions as the ones reflected by quality requirements. While it is certainly not our intention to marginalise existing empirical work in RE, it is reasonable to say that much of the existing body of knowledge in RE remains still a collection of either isolated, loosely connected hypotheses (e.g., empirical insights from a case study in a very specific context) or hypotheses that remain too universal (i.e., completely neglecting the context regardless whether intentional or not). Conclusions are thus either hardly generalizable or easily falsifiable (in a specific context). Examples for existing isolated, yet empirically grounded, observations are:
\begin{itemize}
\item \emph{QRs are mainly elicited by architects}~\cite{Ameller12}
\item \emph{Functional requirements are often labeled as QRs}~\cite{Eckhardt16}
\item \emph{Testing QRs is impossible}~\cite{borg2003bad}
\end{itemize}

Motivated by this overall situation, we want to better understand the extent to which available research statements on quality requirements\footnote{Please note that in our study, we intentionally exclude all non-functional properties that do not address system-specific properties, such as process-related requirements. Hence, we intentionally use the term ``quality requirements'' instead of ``non-functional requirements'' to make explicit that we exclusively focus on non-functional characteristics of a system under consideration. See also \Cref{sec:Background} for further information., as found in academic peer-reviewed and frequently cited publication} are consistent with the perceptions held by practitioners.

In particular, we aim at understanding the extent to which the views and perceptions held by practitioners are corroborated by those of academics. More precisely, we want to understand how well research statements frequently referred to in academic works are perceived by practitioners in their respective context. Questions we opt for answering are:
\begin{itemize}
\item What is the agreement of practitioners with existing research statements about QRs?
\item Which context factors (e.g., industrial sector, company size, experience) influence the agreement of practitioners with research statements about QRs?
\item Can we assign a specific perception of QRs to stereotypical groups of practitioners?
\end{itemize}

Our hope is that an increased understanding of the practitioners' beliefs and views helps us identifying differences, commonalities, and context-dependent grey areas and pinpoint to existing (and regularly cited) statements where further context-dependent research would be useful.

The paper makes the following contributions:
\begin{enumerate}
\item We define a set of 21 research statements about quality requirements from a total of 17 exemplary and commonly cited research papers from the RE research community.
\item We survey practitioners from several application domains and business contexts regarding their agreement with the previously identified statements about quality requirements. The survey results suggest that practitioners hold strong, and diverse opinions, and that some results inspire more passion and dissension than others. 
\item We provide a statistical model that allows evaluating the impact of specific context factors on the perception of research statements. The results of the evaluation show that the perception of some research statements is homogeneous across different development contexts while the perception of others strongly depends on the context.
\item We provide a detailed discussion of the results and contrast them with the original studies from which the statements emerged.
\end{enumerate}

Our intention is not to criticize selected academic manuscripts but to increase our understanding on (1) how much practitioners' views differ with respect to their daily working context, and (2) what we, the research community, can learn from it. Our vision is to contribute to reducing the gap between industrial practice and problems, and academic contributions and solution proposals.

\textbf{Relation to Previous Publications:}
In the past, we have conducted a number of studies in which we investigated the perception~\cite{Eckhardt17NFRvsFR} and use~\cite{Eckhardt16, Eckhardt15} of quality requirements by practitioners. The research questions, results, and the underlying data presented in this article are original in the sense that they have not been addressed in a previous analysis and consequently in a publication. The only commonality between the study at hand and one of our previous publications~\cite{Eckhardt17NFRvsFR} is that the data underlying these studies have been collected using the same questionnaire (but different parts of it). More specifically, we designed a questionnaire on personal experiences to understand whether quality requirements and functional requirements are handled differently in practice. That means, our previous study~\cite{Eckhardt17NFRvsFR} and the study presented in this article share the same subject population but the analyzed questions are completely disjoint. In our 2017 paper~\cite{Eckhardt17NFRvsFR}, we analyzed the answers to questions in questionnaire Sections~3--6, while in the article at hand, we analyze the questions from Section~7 of the questionnaire.
The full questionnaire is part of our additional material package disclosed in our replication package~\cite{Vogelsang18}.

\section{Background}
\label{sec:Background}

In this section, we provide background and related work on QR classifications and on the implications of QRs on software development.

\subsection{QR/NFR Research}
\label{sec:NFRResearch}

\begin{figure}
\centering
  \includegraphics[width=0.8\textwidth]{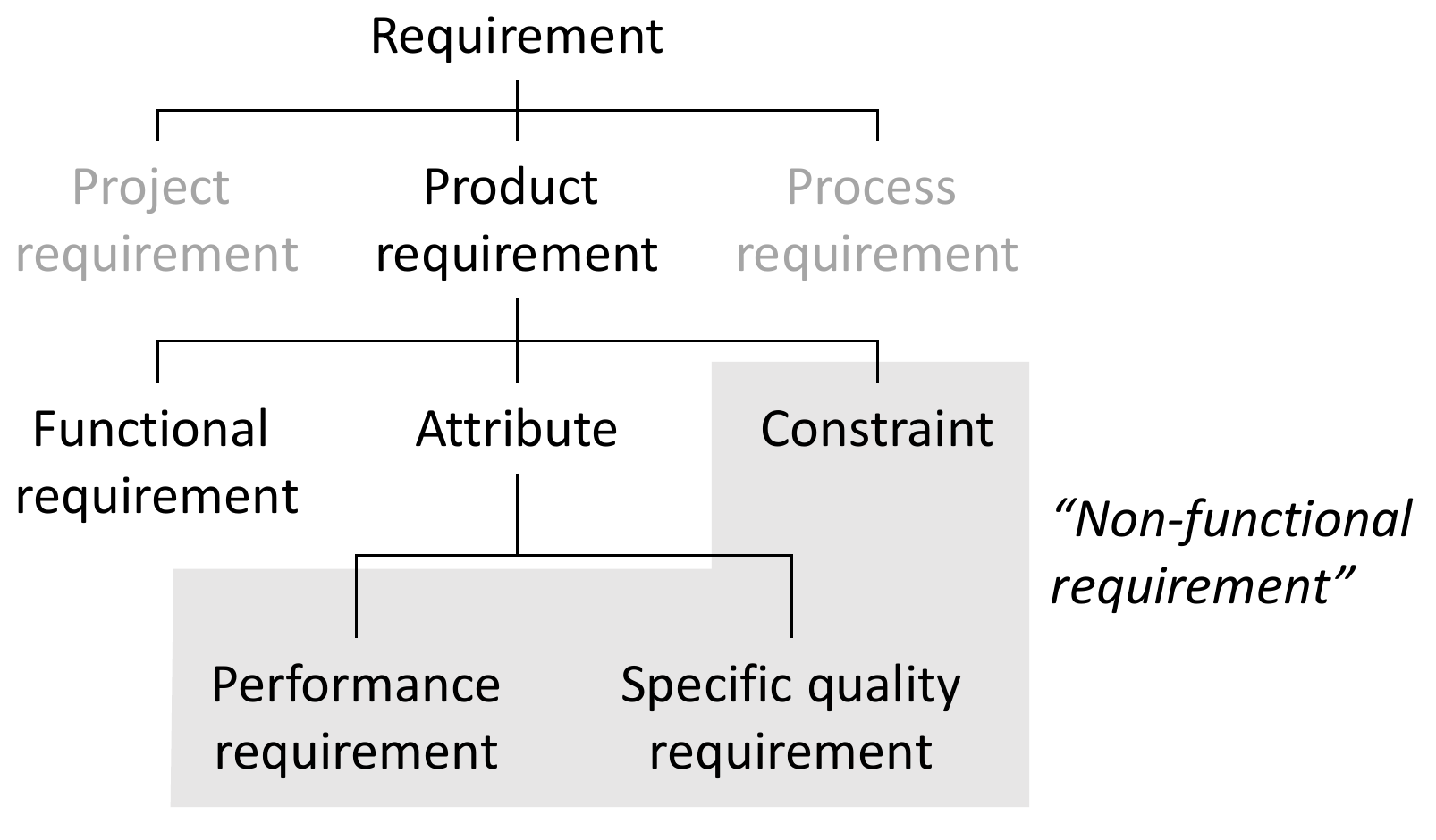}
  \caption{Requirements taxonomy according to Glinz~\cite{Glinz07}.}
  \label{fig:ReqClassification}
\end{figure}

The term {\itshape quality requirement} and the closely related term {\itshape non-functional requirement (NFR)} are subject to constant discussions and even misunderstandings in academia and practice. In fact, there seems to be an agreement in academia that the term ``non-functional requirements'' should be generally avoided when characterizing requirements. In his seminal paper, Glinz~\cite{Glinz07} performs a comprehensive review on the existing definitions related to the term ``non-functional requirement''. He highlights three problems: (1) a definition problem, i.e., NFR definitions have discrepancies in the used terminology and concepts, (2) a classification problem, i.e., the definitions provide very different sub-classifications of NFRs, and (3) a representation problem, i.e., the notion of NFRs is representation-dependent. Similarly, Pohl~\cite{Pohl10} discusses the misleading use of the term ``non-functional'' and argues to use, instead, ``quality requirements'' for product-related NFRs that are not constraints. In this manuscript, we also rely on this distinction and use ``quality requirement'' when particularly referring to product-related non-functional properties. Glinz proposes a requirements classification without even using the term NFR at all. However, he also recognizes the prevalent use of the term NFR and defines it as a ``requirement that is an attribute of or a constraint on a system'' (see \Cref{fig:ReqClassification}).
As stated earlier, in our work, we are interested in product-related (quality) requirements. In the taxonomy of Glinz, our characterization covers \emph{performance requirements} and \emph{specific quality requirements}. However, instead of using Glinz's term \emph{attribute}, we follow the suggestion of Pohl~\cite{Pohl10} and use the term \emph{quality requirements} throughout this manuscript. Broy~\cite{Broy16Rethinking} takes a completely different view towards categorizing requirements. He points out that there is no precise definition of both terms {\itshape functional} and {\itshape non-functional}, and he avoids both in his taxonomy. He argues that a requirements categorization should rather differentiate whether a requirement relates to the system's interface, its internal architecture, its internal state, or whether it prescribes representational aspects. 

Despite the observation that recent academic taxonomies seem to avoid using the term ``non-functional'', too, the term is still widely used in practice and also in scientific papers, mostly in the sense of \emph{everything besides the functional requirements}. Eckhardt~et~al~\cite{Eckhardt16} analyzed 11 requirements specifications from industrial environments with a particular focus on requirements labeled as ``quality'' or ``non-functional''. They found that most requirements specifications separate quality requirements from functional requirements in the documentations. However, when analyzing the quality requirements in detail, they found that many requirements labeled as QR describe system behavior and, thus, could as well be labeled as functional requirements. Jung et al.~\cite{Jung04} performed a study on the adequacy of the quality characteristics defined in the ISO\slash IEC 9126 standard~\cite{ISO9126}. They asked 75 study subjects to rate a given software product in terms of a number of quality sub-characteristics defined in the standard. Afterwards, they clustered the sub-characteristics based on correlation between the given answers. Their results reveal ambiguities in the way that ISO\slash IEC 9126 is structured in terms of characteristics and sub-characteristics. For example, their results imply that four specific sub-characteristics actually measure the same intrinsic concept, which is a mixed concept of maintainability and portability. 

One may argue that discussing requirements categorization is an academic gimmick. However, there is empirical evidence that the categorization influences how requirements are elicited, documented, and validated in practice~\cite{Ameller12,borg2003bad,chung1995dealing,svensson2009quality}. 
In an earlier work~\cite{Eckhardt17NFRvsFR}, we found that the development process for requirements of the two classes strongly differs (e.g., in testing). We obtained these findings based on a survey with practitioners to which we will also refer in the work presented here. We further found that many reasons are based on assumptions rather than on evidence. As a matter of fact, up to now, there does not exist a commonly accepted approach for the QR-specific elicitation, documentation, and analysis~\cite{borg2003bad,svensson2009quality}; QRs are usually described vaguely~\cite{borg2003bad,Ameller12}, remain often not quantified~\cite{svensson2009quality}, and as a result remain difficult to analyze and test~\cite{Ameller12,borg2003bad,svensson2009quality}. Furthermore, QRs are often retrofitted in the development process or pursued in parallel with, but separately from, functional requirements~\cite{chung1995dealing} and, thus, are implicitly managed with little or no consequence analysis~\cite{svensson2009quality}. This limited focus on QRs can result in the long run in high maintenance costs~\cite{svensson2009quality}. 

All these studies indicate, so far, that QRs are not well integrated in practical software development processes and furthermore that several problems are evident with QRs. In this paper, our goal is to analyze the discrepancy between perceptions of QRs in academia and practice.

\subsection{Research about Perception of Research Statements in Software Engineering}
In our work presented here, we want to analyze the perception of practitioners on statements in research about quality requirements. We call those statements research statements. Analyzing the appraisal of practitioners with respect to some normative statements has been targeted by a number of other authors as well.

Devanbu~et~al.~\cite{Devanbu16}, for instance, report on a case study on the prior beliefs of developers at Microsoft, and the relationship of these beliefs to actual empirical data on the projects in which these developers work. Their results suggest that programmers do indeed have very strong beliefs on certain topics, that their beliefs are primarily formed based on personal experience rather than on findings in empirical research, and that beliefs can vary with each project, but do not necessarily correspond with actual evidence in that project. They conclude that more effort should be taken to disseminate empirical findings and that more research is needed on the relation between belief and evidence in software practice.

Rainer et al.~\cite{Rainer03} used content analysis to analyze a group discussion about software process improvement (SPI) between developers within one company and compare the respondents opinion with four research papers on SPI. The main finding from this analysis is that there is an apparent contradiction between developers saying that they want evidence, and what developers will accept as evidence. This main finding is related to issues such as hierarchies of knowledge, the value of empirical evidence to practitioners, local expertise, an incremental approach to improvement that may develop familiarity with those improvements, and differences between developers and managers with regards to their interest in the process. A serious implication follows from the main finding: even if researchers could demonstrate a strong, reliable relationship between software process improvement and organisational performance, there would still be the problem of convincing practitioners that the evidence applies to their particular situation.

The work at hand is inspired by the work of Devanbu et al.~\cite{Devanbu16}, as it is the first work to raise the question of the discrepancy between evidence and belief in software engineering and its empirical investigation. Our work follows their line of reasoning but concentrates on research statements related to QRs. In addition, we perform an in-depth analysis of the influence of several context factors on the practitioners' perceptions in order to uncover research statements that may be valid in certain contexts only. To the best of our knowledge, the paper at hands is the first attempt to empirically investigate research statements about QRs and their perception in practice.

\section{Study Design}
Our overall goal is to analyze the extent to which practitioners agree with statements on quality requirements emerging from academic research. By investigating potential mismatches between academic statements, perceptions of practitioners, and the specific context of practitioners, we hope to be able to identify research gaps for specific contexts, potential misconceptions that raise the need for further investigations, or statements that are only true in certain contexts. Our study consisted of the following steps: 
\begin{enumerate}
\item {\bfseries Identify research statements.}\\
We identified and extracted a set of research statements about quality requirements by browsing the literature from relevant journals and conferences.
\item {\bfseries Collect feedback via a survey.}\\
We integrated all found research statements into an online survey where respondents from industry should state their general agreement with the statements. We sent out the survey to a broad spectrum of practitioners that work with requirements in general.
\item {\bfseries Analyze the data.}\\
Given the responses, we applied a statistical model on each research statement that relates the level of agreement to specific context factors (described in \Cref{subsec:data}). The model allows calculating probabilities of a higher agreement or disagreement given a respondent characterized by some context factors.  
\end{enumerate}

Given these steps, our study approach is a form of \emph{qualitative survey}~\cite{Jansen10}. Qualitative surveys particularly study diversity (not distribution) in a population in a cross-sectional manner. They do not aim at establishing frequencies, means, or other parameters but at determining the diversity of some topic of interest within a given population. This type of survey does not count the number of people with the same characteristic (value of variable) but it establishes the meaningful variation (relevant dimensions and values) within that population. Although this type of survey is coined as \emph{qualitative survey}, it has to be noted that we use a statistical (i.e., quantitative) model to analyze the answers of the survey (see \Cref{sec:analysis}).

\subsection{Identification of Research Statements}
\label{sec:study_design:statements}

In the context of our study, a \emph{research statement} is an assertion about quality requirements that has been stated in a high-quality, scientific publication. Research statements usually correspond to observations that single researchers (the authors of the papers) made in a specific context, even if it is not necessarily made explicit in the papers. Hence, research statements do not necessarily need to be true or commonly accepted by the whole community, but we expect that the research statements considered in our study to show a certain degree of scientific rigor proven by the peer-review process and by a considerable number of citations. 

Our goal is to use a number of research statements as a vehicle to assess the differences and commonalities between the perception of academics and practitioners on the topic of quality requirements.
To this end, we identified research statements on QRs by analyzing existing empirical studies concerning non-functional requirements or quality requirements. This process was not intended to be systematic in the sense of conducting a secondary study such as a systematic literature review, because we did not aim for a complete coverage of all research statements. Instead, we analyzed the relevant literature known to us to extract an exemplary set of research statements covering different RE topic areas. The selection of the research statements was, in that sense, opportunistic. 
Most statements were identified in the introduction and conclusion sections of the considered papers. We further validated and discussed the research statements in the team of authors to strengthen our confidence that the statements are reflected correctly. 
In total, we extracted 21 research statements about quality requirements. Please further note that although the extracted statements are based on publications, they are not necessarily taken verbatim and they do not necessarily reflect the authors' opinions. 
Finally, at a later phase of the study, we compared the identified research statements also against recent publications that we found in major venues on requirements engineering and software engineering (in particular, the conferences RE, REFSQ, ESEM, and ICSE) in the time between 2010 and 2016. More specifically, we analyzed all publications categorized as concerning quality requirements in another literature study~\cite{Franch17} and looked for statements that either support or oppose research statements from our initial list of research statements.
\Cref{tbl:researchStatementsList} summarizes the research statements we consider for this study and adds further publications supporting or opposing the statements. For a better overview, we clustered the research statements according to the different RE activities.

\begin{table}
\renewcommand{\arraystretch}{1.3}
\caption{Identified research statements considered in our study with publications supporting (pro) and opposing (con) the statements.}
\label{tbl:researchStatementsList}
\centering
\begin{tabularx}{\textwidth}{@{}p{0.2cm}p{0.4cm}Xp{1.1cm}p{1.1cm}@{}}
\toprule
&{\bfseries Id}&{\bfseries Research Statement}&{\bfseries Pro} &{\bfseries Con}\\ 
\midrule          
\parbox[t]{0mm}{\multirow{8}{*}{\rotatebox[origin=c]{90}{\bfseries General}}} 
&{\bfseries G1} &The application domain strongly influences the relevance of individual types of QRs. & \cite{Eckhardt16,Rahimi14} & \\
&{\bfseries G2} &Many QRs describe functional aspects of a system. & \cite{Eckhardt16} & \cite{Feng14}\\
&{\bfseries G3} &QRs are sometimes ignored. & \cite{borg2003bad} & \\
&{\bfseries G4} &Architects do not share a common terminology for types of QRs.& \cite{Ameller12,Daneva14,Mahmoud15} & \cite{Daneva13} \\
&{\bfseries G5} &Only few QRs deal with architectural aspects. & \cite{Eckhardt16} & \cite{Clements10,Poort12}\\
\midrule
\parbox[t]{0mm}{\multirow{6}{*}{\rotatebox[origin=c]{90}{\bfseries Elicitation}}} 
&{\bfseries E1} &In requirements elicitation, the focus is on FRs, not on QRs.& \cite{borg2003bad,Ameller12,Svensson11,Rahimi14,Felderer14,Mahmoud15} &\\
&{\bfseries E2} &Many QRs remain undiscovered.& \cite{borg2003bad,Rahimi14,Mahmoud15}&\\
&{\bfseries E3} &QRs are mainly elicited by architects.& \cite{Ameller12,Daneva13}&\cite{Clements10,Daneva14}\\
\midrule
\parbox[t]{0mm}{\multirow{10}{*}{\rotatebox[origin=c]{90}{\bfseries Specification}}} 
&{\bfseries S1} &QRs are often not documented. & \cite{Ameller12,Rahimi14} & \cite{Daneva13,Daneva14}\\
&{\bfseries S2} &The documentation of QRs is not always precise.& \cite{Ameller12,svensson2009quality,Fotrousi14,Feng14} &\\
&{\bfseries S3} &QRs are often described in too vague terms.& \cite{borg2003bad,Feng14} &\\
&{\bfseries S4} &The documentation of QRs usually becomes desynchronized.& \cite{Ameller12,Mirakhorli12,Mahmoud15} &\\
&{\bfseries S5} &Functional requirements are often labeled as QRs.& \cite{Eckhardt16,Ernst10,Rahimi14} &\\
&{\bfseries S6} &QRs are often specified by referencing a standard or a legislative text.& \cite{Eckhardt16,Daneva13} &\\

\midrule
\parbox[t]{0mm}{\multirow{8}{*}{\rotatebox[origin=c]{90}{\bfseries Testing}}} 
&{\bfseries T1} &Only few types of QRs are validated at the end of the project.& \cite{Ameller12} & \cite{Daneva13}\\
&{\bfseries T2} &QRs are satisfied at the end of the project.& \cite{Ameller12,Ernst10,Nguyen12} &\\
&{\bfseries T3} &Most QR types are difficult to test properly.& \cite{borg2003bad,svensson2009quality,Fotrousi14,Mahmoud15} &\\
&{\bfseries T4} &Testing QRs is time-consuming.& \cite{borg2003bad} &\\
&{\bfseries T5} &Testing QRs is impossible.& \cite{borg2003bad} &\\

\midrule
\parbox[t]{0mm}{\multirow{3}{*}{\rotatebox[origin=c]{90}{\bfseries Mgmt.}}} 
&{\bfseries M1} &QRs are often not sufficiently prioritized.& \cite{borg2003bad,Svensson11}&\cite{Daneva13}\\
&{\bfseries M2} &Software architects do not use a specific tool for QR management.& \cite{Ameller12} &\\

\bottomrule
\end{tabularx}
\end{table}

\subsection{Subject Selection}
With our study, we targeted at practitioners who work with requirements. This includes practitioners who write requirements (e.g., \emph{requirements engineers}), practitioners whose work is based on requirements (e.g., \emph{developers} or \emph{testers}), and practitioners who manage projects or requirements.
For inviting practitioners to participate, we did not select a specific closed group of practitioners but, instead, contacted as many practitioners as possible via the authors' personal contacts from previous collaborations, via public mailing lists such as \emph{RE-online}, and via social networks. 
Our survey was further conducted anonymously. Since we were not able to exactly control who is answering the survey, it was especially important to follow Kitchenham and Pfleeger's~\cite{Kitchenham08} advice on the need to understand whether the respondents had enough knowledge to answer the questions in an appropriate manner. For this, we excluded data from respondents who answered that they do not use requirements specifications at all or respondents who stated that they did not know how requirements are handled in their company. 
We offered respondents the chance to leave an email address if they were interested in the results of the survey.

\subsection{Data Collection} \label{subsec:data} 
We integrated the survey questions of this paper into a larger survey about general QR practices using the Enterprise Feedback Suite EFS Survey from Questback. We published the results of this larger study in a previous paper~\cite{Eckhardt17NFRvsFR}. However, in the previously published paper, we did not report on the results or discuss any of the questions from this study. We started our data collection on February 4th, 2016 and closed the survey on February 22nd, 2016. In the following, we introduce the main elements of our instrument used. The full instrument is part of our additional material package disclosed in our replication package~\cite{Vogelsang18}. 

\subsubsection{Subject Matter Clarification}

In the survey questionnaire, we wanted to ensure that all respondents have a similar understanding about the subject matter. Therefore, we first introduced a common terminology (``With NFRs, we refer to those requirements that address quality characteristics of a product or system (like availability or performance)'') and further narrowed down the scope via specific questions on a set of specific quality characteristics. In the survey, we intentionally used the term ``non-functional requirements'' when referring to ``quality requirements'' (excluding process and project requirements as well as constraints), because the term NFR is widely used in practice (see also \Cref{sec:NFRResearch}). 
In the remainder of this paper, we will exclusively use the term ``quality requirements''.

\subsubsection{Demographics and Context Factors} 
We collected demographic data from the respondents to be able to interpret and triangulate the data with respect to different contexts of the respondents. The elicitation of demographic data included the following context factors:
\begin{itemize}
  \item \textbf{Role in project:} Free text answers that we afterwards categorized to the project roles \emph{manager}, \emph{requirements engineer}, \emph{architect}, \emph{test engineer}, \emph{developer}, and \emph{other}.
  \item \textbf{Experience:} Choice between less than 3 years of experience in dealing with requirements (novice) and more than 3 years of experience (senior).
  \item \textbf{Company size:} Choice between less than 250 employees (small company), between 250 and 2,000 employees (medium company), and more than 2,000 employees (large company).
  \item \textbf{Typical project size:} Choice between less than 10 employees (small projects), between 10 and 50 employees (medium projects), and more than 50 employees (large projects).
  \item \textbf{Geographical team distribution:} Choice between \emph{all team members in one location}, \emph{team members distributed over several locations in one country}, and \emph{team members distributed over several locations in several countries}. 
  \item \textbf{Development process paradigm:} Choice between \emph{rather agile}, \emph{mixed}, \emph{rather plan-driven}.
  \item \textbf{Industrial sector:} Free text answers that we afterwards categorized to the sectors \emph{telecommunication}, \emph{automotive}, \emph{automation}, \emph{avionics}, \emph{finance}, \emph{healthcare}, \emph{public}, and \emph{other}.
  \item \textbf{System type:} Choice between \emph{embedded systems}, \emph{business information systems}, \emph{consumer software}, and \emph{hybrid systems}.
  \item \textbf{Role of requirements specifications in the company:} Choice between \emph{create and use for in-house development}, \emph{create and an external company is responsible for the development}, \emph{use as a subcontractor for e.g. development or testing}, and \emph{don't use}.
  \item \textbf{Documentation of QRs:} Choice between \emph{yes} (QRs are documented) and \emph{no} (QRs are not documented).
 
\end{itemize}

To better understand the participant's focus and project context, we additionally asked the respondents to evaluate the importance of different types of QRs in their projects. 
The respondents were asked to assess the importance of quality factors taken from ISO\slash IEC~25010~\cite{iso25010} for their typical projects on a 5-point Likert scale with the values ``very important'', ``important'', ``moderately important'', ``slightly important'', ``not important''. We added another value ``Don't know'' that allows respondents to skip the answer if they can or do not want to answer the question.
The quality factors that we asked for were \emph{functional suitability}, \emph{performance\slash efficiency}, \emph{compatibility}, \emph{usability}, \emph{reliability}, \emph{security}, \emph{maintainability}, and \emph{portability}. 

\subsubsection{Research Statement Agreement}
Finally, we presented all research statements to the respondents and asked them to state their agreement with the research statement: ``Please consider your experiences: To which degree do you agree with the following statements?'' The research statements were presented in an arbitrary order that was different for all respondents.
The respondents could express their agreement on a 5-point Likert scale with the values ``strongly agree'', ``agree'', ``neither agree nor disagree'', ``disagree'', ``strongly disagree''. We added another value ``Don't know'' to allow respondents to express that they have no opinion about or cannot answer the question.
The last category was included to address the diverse background of respondents---not all respondents will understand all research statements. 

\subsection{Data Analysis}
\label{sec:analysis}
To asses the relation between the agreement with the research statements and the context factors, we set up a regression model that we applied to all the research statements. For this purpose, the collected data has to be coded appropriately. The following representation focuses on one research statement only. Let the data be given by $(y_i,\boldsymbol{x}_i),\; i=1,\hdots,n$, where $y_i$ is the response to the research statement of respondent $i$ from 1 (strongly agree) to 5 (strongly disagree), $\boldsymbol{x}_i$ is the vector of context factors, and $n$ is the number of respondents. Each context factor contained in $\boldsymbol{x}_i$ is coded according to its scale. For the binary context factors, we use usual dummy variables (i.e., auxiliary variables taking the value 0 or 1). They are \textit{Experience [0: novice, 1: senior]} and \textit{Documentation of QRs [0: no, 1:yes]}. For the categorical context factors, we use an effect coding scheme, where the effect of the last category is fixed, respectively. The coding of two examples with three categories is \textit{Company size [(1,0): small, (0,1): medium, (-1,-1): large]} and \textit{Development process paradigm [(1,0): rather agile, (0,1): mixed, (-1,-1): plan-driven]}. For the other categorical context factors it works in the same way. For simplicity, the answers on the importance of quality factors were treated as numeric variables in the model. 

In the regression model, the response to the research statement is treated as an ordinal variable, that is, one explicitly uses the ordering of the variable. An important class of ordinal regression models is the class of cumulative models~\cite{Agresti:2010},~\cite{LiuAgr:2005}. The most prominent one is the proportional odds model, which is applied here. The basic form of the model is given by 
\begin{equation} 
\label{eq:model}
\log \left(\frac{P(y_i \le r|\boldsymbol{x}_i)}{P(y_i > r|\boldsymbol{x}_i)}\right) = \theta_{r} + \boldsymbol{x}_i^\top\boldsymbol{\beta}, \quad r=1,\dots,4\,, 
\end{equation}
where $\theta_{r}$ are category specific threshold parameters and $\boldsymbol{\beta}$ is the vector of regression coefficients (estimated impact of the context factors) that is independent of the category $r$. Consistent estimates of the regression coefficients are obtained by maximizing the log-likelihood function by means of the Fisher scoring algorithm~\cite{FahTut:2001}. Estimation was carried out by the statistical software \texttt{R}~\cite{R} using function \texttt{vglm()} of the add-on package \texttt{VGAM}~\cite{yee2010vgam},~\cite{VGAM2014}.  

\subsubsection{Interpretation of the Parameters} 
\label{sec:regression_coefficient}
In \Cref{eq:model} the threshold parameters $\theta_{r}$ define the general preference for the categories of a research statement (the general level of agreement) and the parameters $\boldsymbol{\beta}$ determine the shifting of the agreement distribution by the context factors. In detail, let us consider two respondents with context factors $\boldsymbol{x}_1$ and $\boldsymbol{x}_2$ and the corresponding cumulative odds $\gamma(r|\boldsymbol{x}_1)={P(y \le r|\boldsymbol{x}_1)}/{P(y > r|\boldsymbol{x}_1)}$ and $\gamma(r|\boldsymbol{x}_2)={P(y \le r|\boldsymbol{x}_2)}/{P(y > r|\boldsymbol{x}_2)}$. Simple derivation shows that the proportion of the cumulative odds for the two respondents is given by
\[
\frac{\gamma(r|\boldsymbol{x}_1)}{\gamma(r|\boldsymbol{x}_2)} = \exp\left((\boldsymbol{x}_1-\boldsymbol{x}_2)^\top\boldsymbol{\beta}\right),
\]
and therefore does not depend on the category $r$. Consequently, the interpretation of parameters does not depend on the category. More concise, $\exp(\beta_j)$ represents the factor by which all the cumulative odds ${P(y \le r|\boldsymbol{x})}/{P(y > r|\boldsymbol{x})}$ change if context factor $x_j$ increases by one unit, while all the other context factors remain constant. For binary context factors, $\exp(\beta_j)$ corresponds to the difference between the two groups, for example, \textit{senior} compared to \textit{novice}. For categorical variables with effect coding, $\exp(\beta_{jk})$ corresponds to the difference between group $k$ and a ``middle group'', respectively. Accordingly, each estimate has to be interpreted as the effect of one group compared to a ``middle effect''. 

If $\exp(\beta_j)$ is positive, the distribution is shifted to the left, which means more agreement. On the other hand, if $\exp(\beta_j)$ is negative, the distribution is shifted to the right, which corresponds to less agreement. 

\subsubsection{Hypothesis Tests}
Standard errors to examine the significance of each context factor can be obtained by asymptotic theory. For a detailed description on inference techniques, see~\cite{FahTut:2001} and~\cite{Tutz:2012}. It is well known that the covariance of the estimates $\hat{\boldsymbol{\beta}}$ is asymptotically given by the expected Fisher matrix~\cite{FahTut:2001}. This allows us to use Wald tests~\cite{Wald43} for the null hypotheses $H_0:\beta_j=0$ against $H_1:\beta_j\neq 0$. 
A Wald test is a classical approach to hypothesis testing of coefficients in a regression model. The test provides a \emph{p}-value and a test statistic for differences in the coefficients for different context factors. Small sample sizes for certain context factors may lead to larger confidence intervals when the Wald test is applied and thus increase the likelihood of \emph{p}-values above the significance level.
In \Cref{sec:study_results}, we give the results of the Wald tests based on significance (type~\uppercase\expandafter{\romannumeral 1\relax}~error) level $\alpha=0.05$. 

\subsection{Validity Procedures}
\label{sec:study_design:v-procedures}
To strengthen the confidence in our results, we performed a few validity procedures in advance.
To ensure that our respondents are really practitioners we explicitly stated that the survey is aimed at addressing practitioners in the introductory text of the survey. In addition, we removed those respondents from the population that stated that they do not deal with requirements.

To lower the threat of biased answers of our respondents, we conducted the survey anonymously and asked additional questions to characterize the context of the respondents. 

To lower the threat that respondents misunderstand or misinterpret particular questions in the questionnaire, we conducted a pilot phase with three practitioners in which we tested and improved the instrument used, but also to evaluate the envisioned data analysis techniques based on the pilot data (which we deleted again prior to starting the survey). 

\section{Study Results} \label{sec:study_results}
In total, 283 people followed the link to our online survey, 172 started the survey (61\%), and 109 completed it (39\%). From these 109 respondents, we excluded 6 as they matched our exclusion criteria (no knowledge about how requirements are handled in their company). 
The survey answers are also available as .csv file in our additional replication package~\cite{Vogelsang18}.

After reporting the demographics, we provide the following results: 
\begin{enumerate}
\item An overview of the general agreement for each research statement (\Cref{sec:res_agreement} and especially \Cref{fig:answerDistribution}). 
\item An overview of the significant context factors for each research statement (\Cref{sec:res_impact} and especially \Cref{tbl:researchStatements}). 
\item A visual overview of the significant impacts on research statements for each context factor (\Cref{sec:res_contextFactors} and especially \Cref{fig:contextfactoranalysis1,fig:contextfactoranalysis2}).
\item An aggregated view of the research statements with respect to consensus and mean agreement (\Cref{sec:res_summary} and especially \Cref{fig:sdmean}). 
\end{enumerate}

\subsection{Demographics}
\begin{figure}
   \centering
     \subfloat[Roles]{
	      \includegraphics[width=0.48\textwidth]{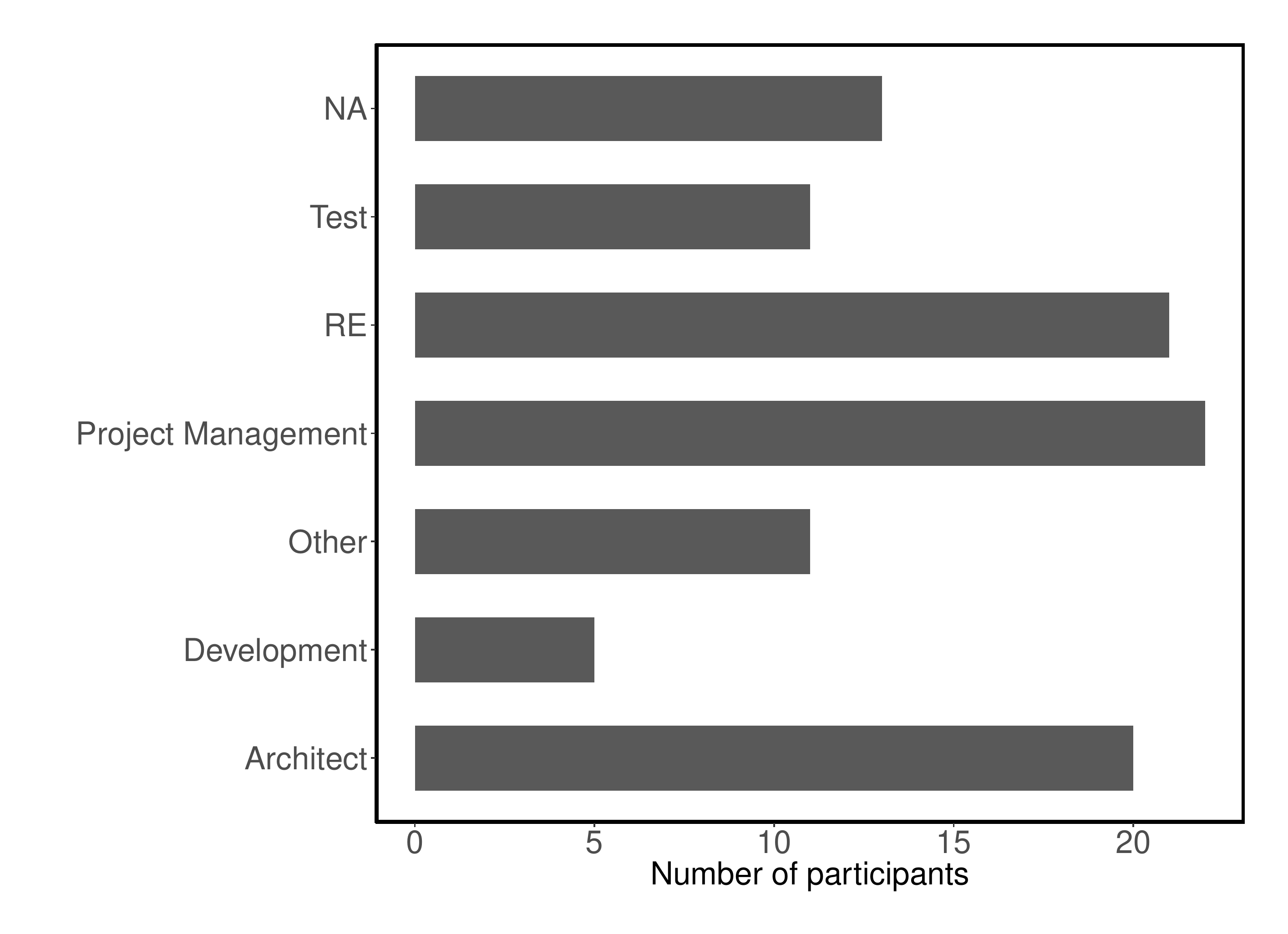}
	      \label{fig:roles}
      }
      \subfloat[Sector]{
	      \includegraphics[width=0.48\textwidth]{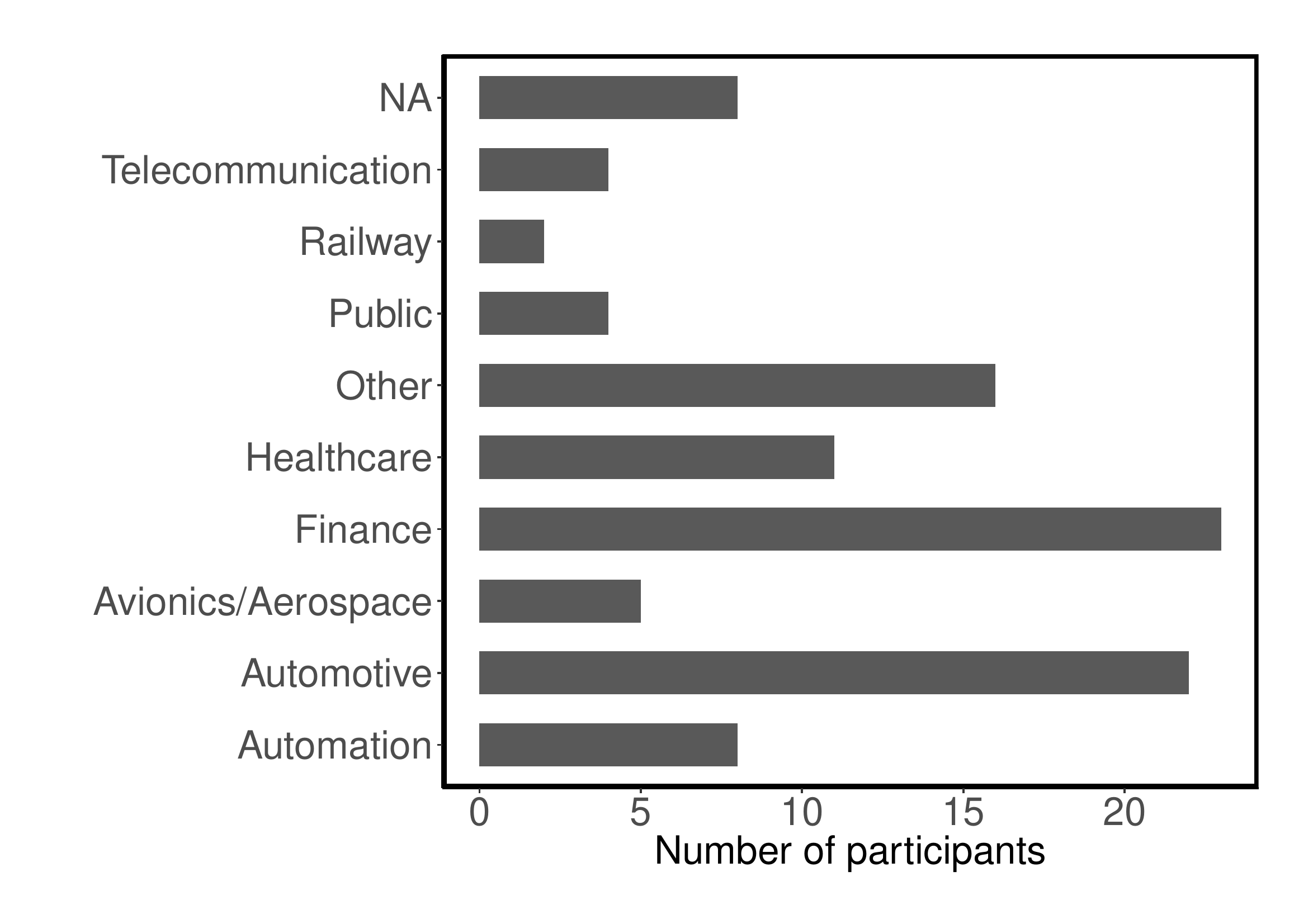}
	      \label{fig:sector}
      }
      
      \subfloat[Company Size]{
	      \includegraphics[width=0.48\textwidth]{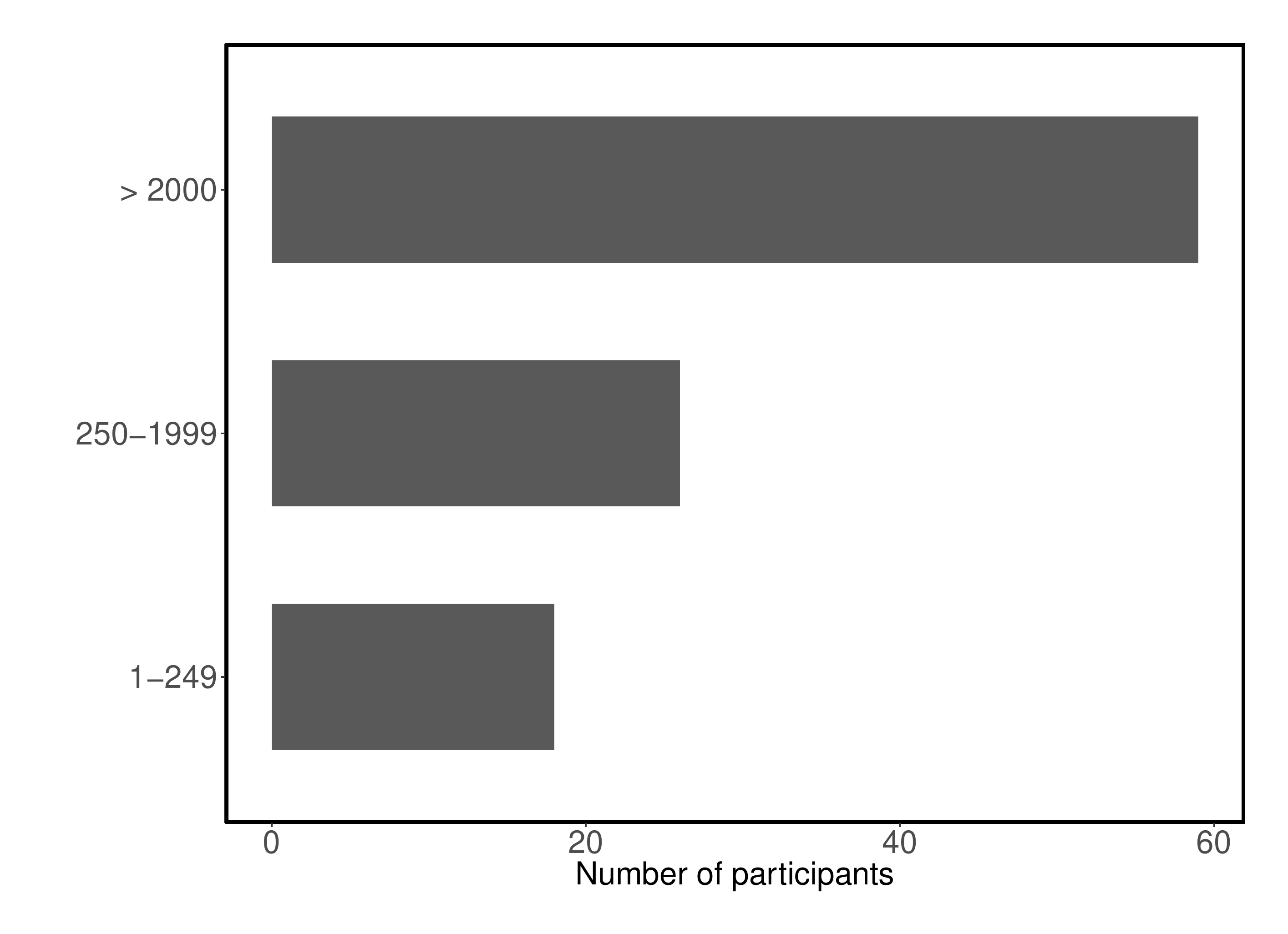}
	      \label{fig:cSize}
      }
      \subfloat[Project Size]{
	      \includegraphics[width=0.48\textwidth]{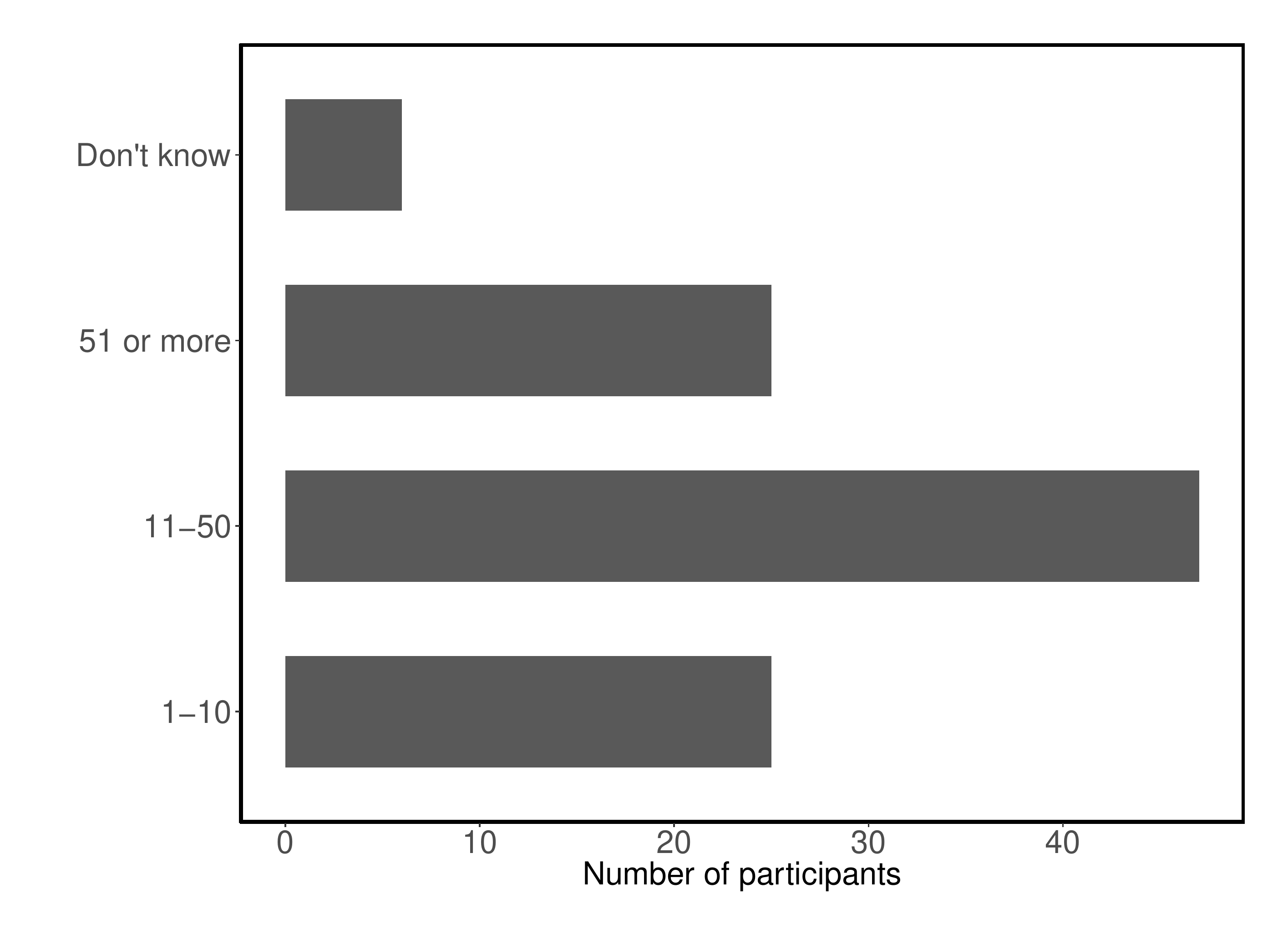}
	      \label{fig:pSize}
      }
      
      \subfloat[Process Paradigm]{
	      \includegraphics[width=0.48\textwidth]{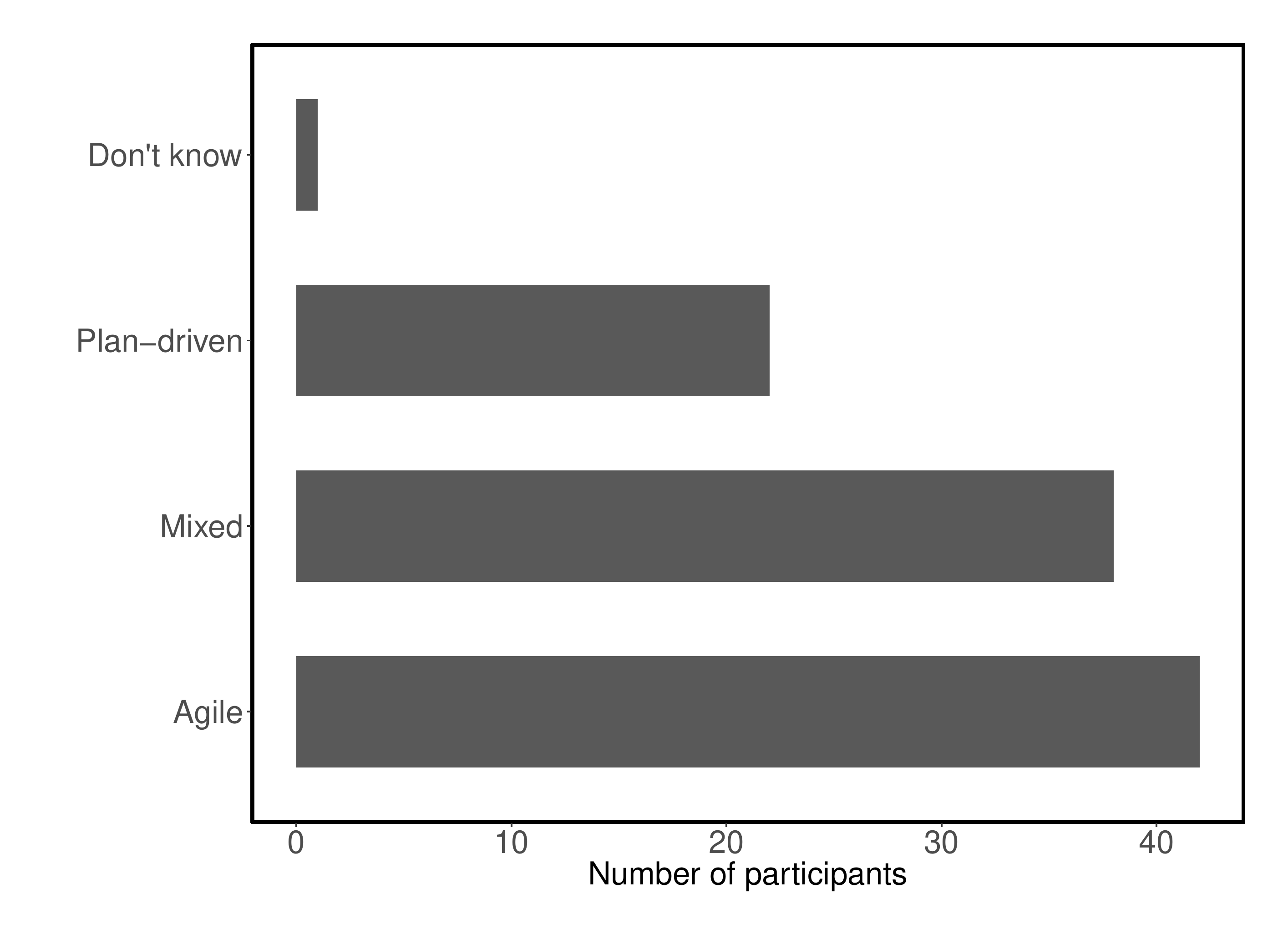}
	      \label{fig:procpara}
      }
      \subfloat[Experience]{
	      \includegraphics[width=0.48\textwidth]{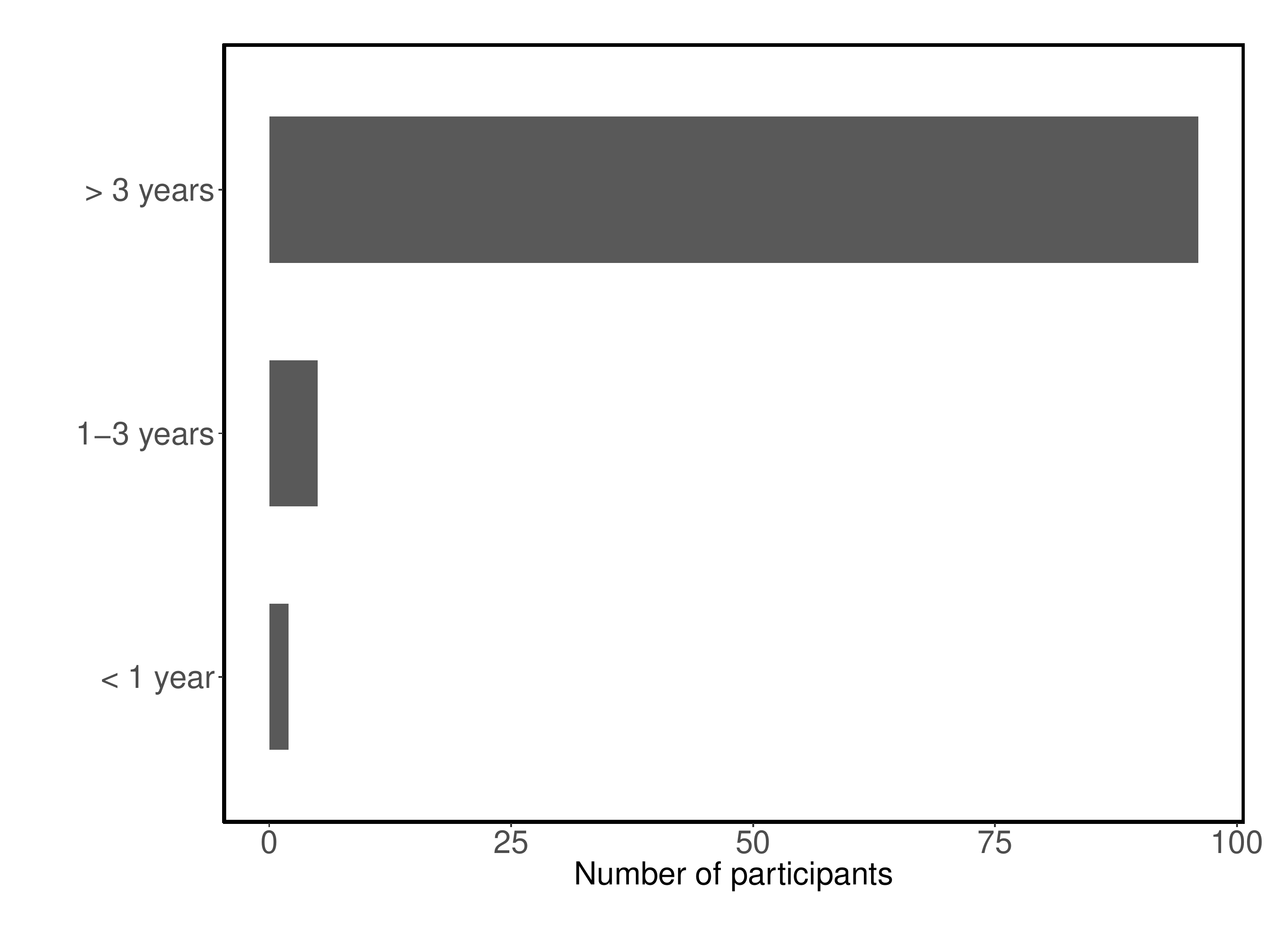}
	      \label{fig:experience}
      }
      
      \subfloat[Type of System]{
	      \includegraphics[width=0.48\textwidth]{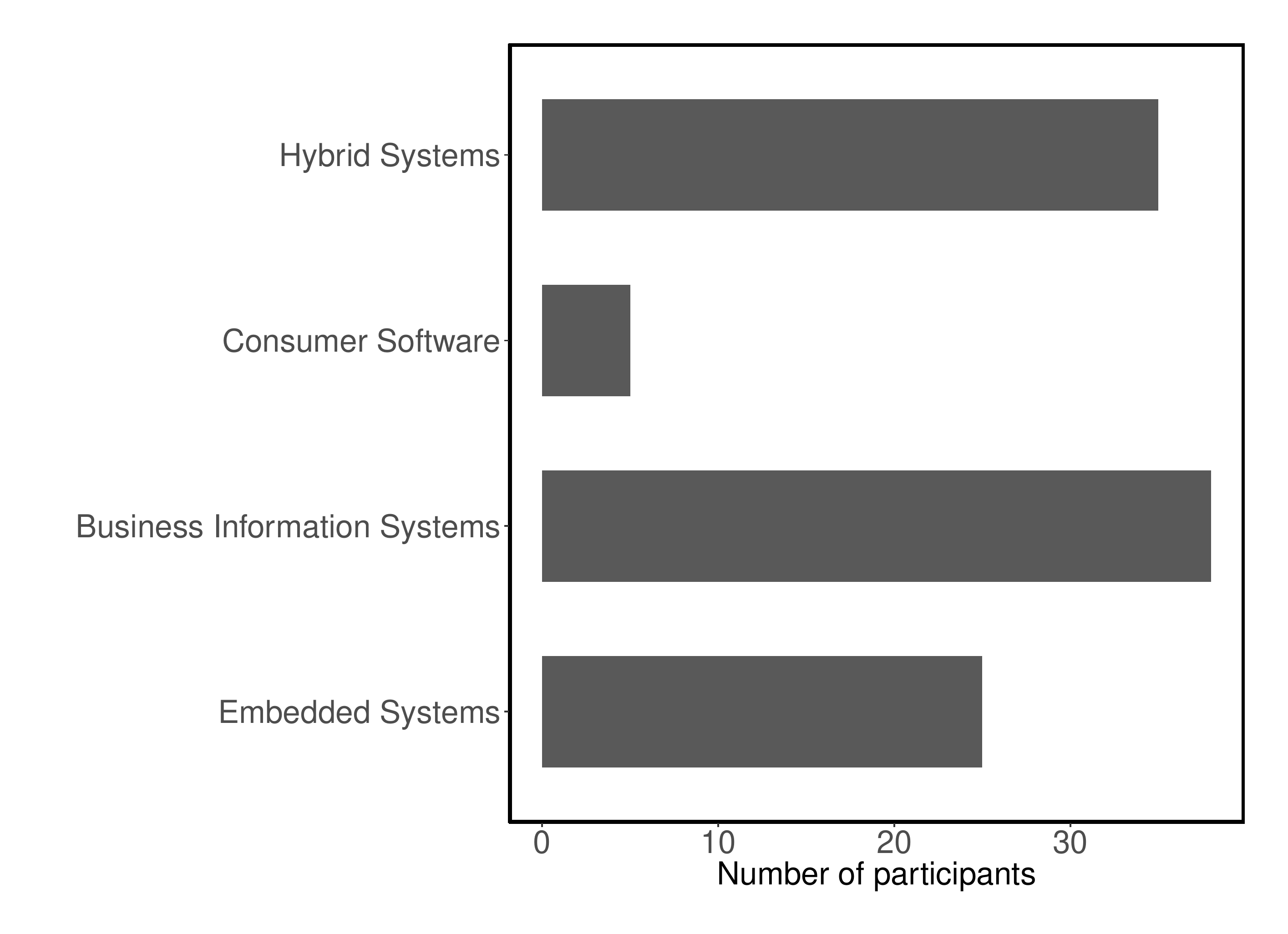}
	      \label{fig:systemType}
      }
      \subfloat[Geographical Distribution]{
	      \includegraphics[width=0.48\textwidth]{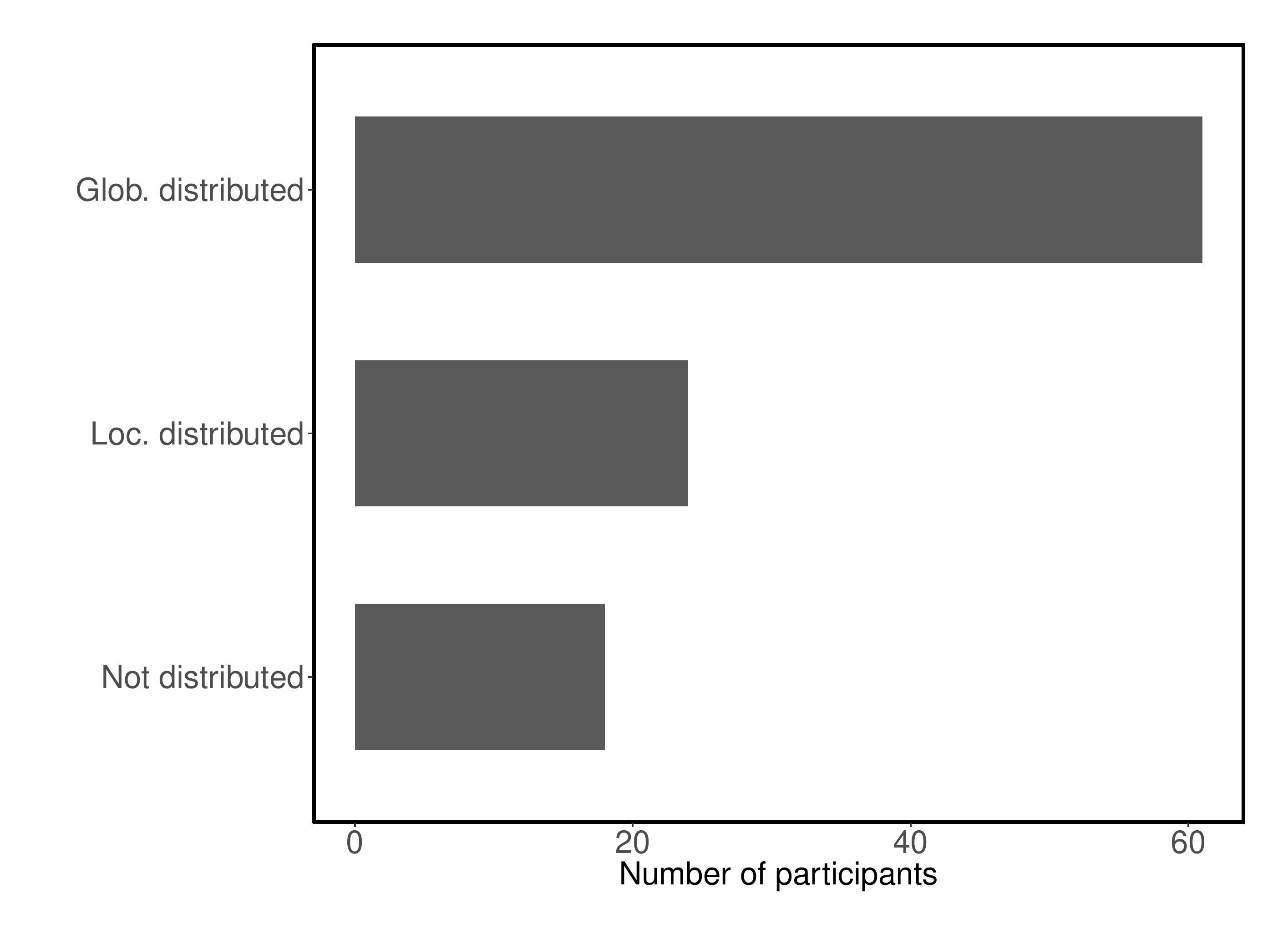}
	      \label{fig:geoDist}
      }
   \caption{Overview of study population}
  \label{fig:population}   
\end{figure}

\Cref{fig:population} shows an overview over our study population. The figures show that our respondents cover a wide spectrum of context factors.

\subsection{General Agreement with Research Statements}
\label{sec:res_agreement}

\begin{figure}
\centering
  \includegraphics[width=\textwidth]{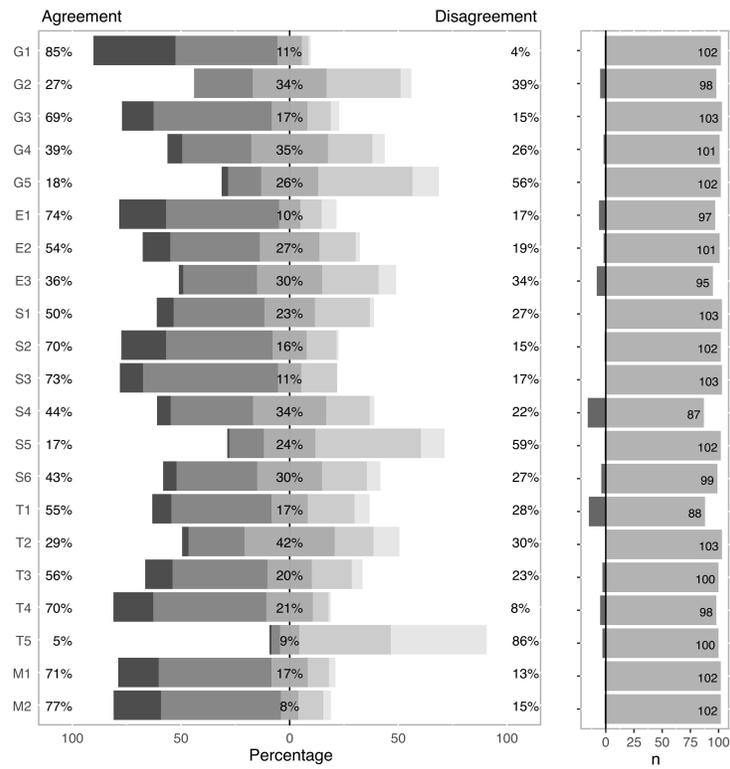}
   \caption{Distribution of answers for all research statements. Left: from strongly agree (dark gray) to strongly disagree (light gray). Additionally, the total percentage of agreement (and disagreement) is shown and the left (right) side of the plot. Right: the number of total answers in gray and the number of \emph{don't know} answers in dark gray.}
  \label{fig:answerDistribution}
\end{figure}

\Cref{fig:answerDistribution} provides an overview of the answer distribution for each research statement. For each statement, on the left side, distribution from strongly agree over neutral (centered) to strongly disagree is shown from dark gray to light gray. On the right side, the total number of respondents who answered the question is shown (the dark gray bar indicates the percentage of \emph{don't know} answers). On average, 99 (median of 101) respondents answered each question, with a minimum of 87 (S4) and maximum of 103 (G3, S1, S3, T2).

\subsection{Impact of Context Factors to the Level of Agreement}
\label{sec:res_impact}
\Cref{tbl:researchStatements} shows an overview of the significant context factors and the corresponding estimated regression coefficients for each research statement. The table only contains context factors for which the regression model indicates a statistically significant impact to the level of agreement (p-value: $p<0.05$). 
A row in this table can be read as follows: For the research statement stated in the first column (e.g., G1), respondents belonging to one of the groups listed in the second column tend to agree more with this statement (e.g., respondents using an agile process paradigm,  respondents who stated that maintainability is important, and respondents working in medium size projects). Respondents belonging to one of the groups listed in the right-hand side column tend to disagree more with this statement (e.g., respondents working in small projects or respondents who stated that portability is important). 
The change factor $\exp(\beta_j)$ provided in brackets after each context factor represents the factor by which the probability to agree or disagree more changes if this context factor changes (see also \Cref{sec:regression_coefficient}).

\begin{table}
\footnotesize
\caption{Relation between context factors and level of agreement for each research statement. The change factor (given in brackets) indicates the factor by which the agreement\slash disagreement changes if this context factor is changed by one unit, while all the others remain constant.
}
\label{tbl:researchStatements}
\centering
\begin{tabularx}{1.05\textwidth}{@{}lXX@{}}
\toprule
{\bfseries RS}&{\bfseries Tendency to agree more}&{\bfseries Tendency to disagree more}\\ 
\midrule          
% The application domain strongly influences the relevance of individual types of QRs.~\cite{Eckhardt16}
{\bfseries G1}&	
Agile process paradigm (5.1), Maintainability important (4.0),  Medium projects (3.1) & 
Portability important (2.1), Small projects (6.8)\\
%Many QRs describe functional aspects of a system.~\cite{Eckhardt16}
{\bfseries G2}	& - & - \\
% QRs are sometimes ignored.~\cite{borg2003bad}
{\bfseries G3}& 
Consumer SW (53.1), Automation sector (15.9),  Medium projects (2.8) & 
Compatibility important (2.4), Usability important (3.1),  Small projects (5.6), Embedded Systems (13.7) \\
% Architects do not share a common terminology for types of QRs.
{\bfseries G4}&	
Healthcare sector (11.5), Non-distributed project (5.2), & 
Small projects (4.6)\\
% Only few QRs deal with architectural aspects.~\cite{Eckhardt16}	
{\bfseries G5}& 
Automotive sector (15.2), Testers (8.3), Managers (3.9),  Nat. distributed projects (3.3), Compatibility important (2.3) & 
Requirements engineers (4.5), Small companies (4.5)\\
\midrule

%In requirements elicitation, the focus is on FRs, not on QRs.~\cite{borg2003bad,Ameller12}                 
{\bfseries E1} & 
Performance\slash Efficiency important (3.7) & 
Managers (5.3), Nat. distributed projects (6.8),  Railway sector (75.8)\\
% Many QRs remain undiscovered.	
{\bfseries E2}&
Healthcare sector (9.0), Small companies (7.7),  Performance\slash Efficiency important (2.8), Medium projects (2.8),  Maintainability important (2.8)& 
Funct. suitability important (2.7), Reliability important (3.4),  Small projects (6.6), Testers (6.9), Avionics sector (26.7),   Railway sector (41.2)\\
% QRs are mainly elicited by architects.	
{\bfseries E3}&
Architects (24.9), Healthcare sector (7.3), Automotive sector (6.6), Inhouse dev. (3.2), Performance\slash Efficiency important (3.2)& 
Requirements engineers (4.0)\\
\midrule
% QRs are often not documented.~\cite{Ameller12}
{\bfseries S1}& 
Architects (7.6)& 
Medium companies (3.5), Embedded systems (6.8),  Railway sector (70.9)\\
% The documentation of QRs is not always precise.~\cite{Ameller12,svensson2009quality}	
{\bfseries S2}& - & - \\
% QRs are often described in too vague terms.~\cite{borg2003bad}	
{\bfseries S3}&
Automation sector (15.8) & 
Automotive sector (6.2), Documented QRs (29.0)\\
% The documentation of QRs usually becomes desynchronized.~\cite{Ameller12}	
{\bfseries S4}&	
Developers (104.2), Telecommunication sector (24.8), Performance\slash Efficiency important (3.3) &
Portability important (2.2),   Reliability important (5.0),  Business information systems (5.7), Requirements engineers (6.8),  Funct. suitability important (8.7), Automation sector (22.0),  Railway sector (69.6)\\
% Functional requirements are often labeled as QRs.~\cite{Eckhardt16}
{\bfseries S5}	&
Agile process paradigm (6.7), Embedded systems (5.7), Architects (4.4), Mixed process paradigm (1.9)& Portability important (2.0), Plan-driven process paradigm (4.4),  Business information systems (6.6)\\
% QRs are often specified by referencing a standard or a legislative text.~\cite{Eckhardt16}
{\bfseries S6}&Performance\slash Efficiency important (3.2) &-\\
\midrule
%Only few types of QRs are validated at the end of the project.~\cite{Ameller12}
{\bfseries T1}&
Healthcare sector (12.4), Automotive sector (12.2),  Business information systems (4.6)& 
Portability important (2.3), Consumer SW (66.7),  Railway sector (261.6)\\
%QRs are satisfied at the end of the project.~\cite{Ameller12}
{\bfseries T2}&
- & 
Telecommunication sector (9.3)\\
% Most QR types are difficult to test properly.~\cite{borg2003bad,svensson2009quality}	
{\bfseries T3}& 
Automotive sector (6.6), Managers (6.4),  Business information system (5.3), Agile process paradigm (3.5) & 
Non-distributed projects (6.9), Finance sector (14.5), Seniors (17.7)\\
% Testing QRs is time consuming.~\cite{borg2003bad}	
{\bfseries T4}&
Avionics sector (59.4), Agile process paradigm (8.3),  Architects (8.3) & 
Portability important (2.6), Testers (6.8) \\
% Testing QRs is impossible.~\cite{borg2003bad}	
{\bfseries T5}&
Non-distributed projects (6.6), Inhouse dev. (3.1),  Security important (3.1) & 
Small companies (7.6), Seniors (27.6)\\
\midrule
% QRs are often not sufficiently prioritized.~\cite{borg2003bad}
{\bfseries M1}&	
Architects (5.7)& 
Usability important (2.5), Testers (13.0), Seniors (17.8)\\
% Software architects do not use a specific tool for QR management.
{\bfseries M2}& 
Consumer SW (120.7), Reliability important (14.5),  Medium companies (8.0) & 
Small companies (96.9)\\
\bottomrule
\end{tabularx}
\end{table}

\subsection{Context Factor Analysis}
\label{sec:res_contextFactors}
\begin{figure}
   \centering
      \subfloat[Role]{
	      \includegraphics[width=0.4\textwidth]{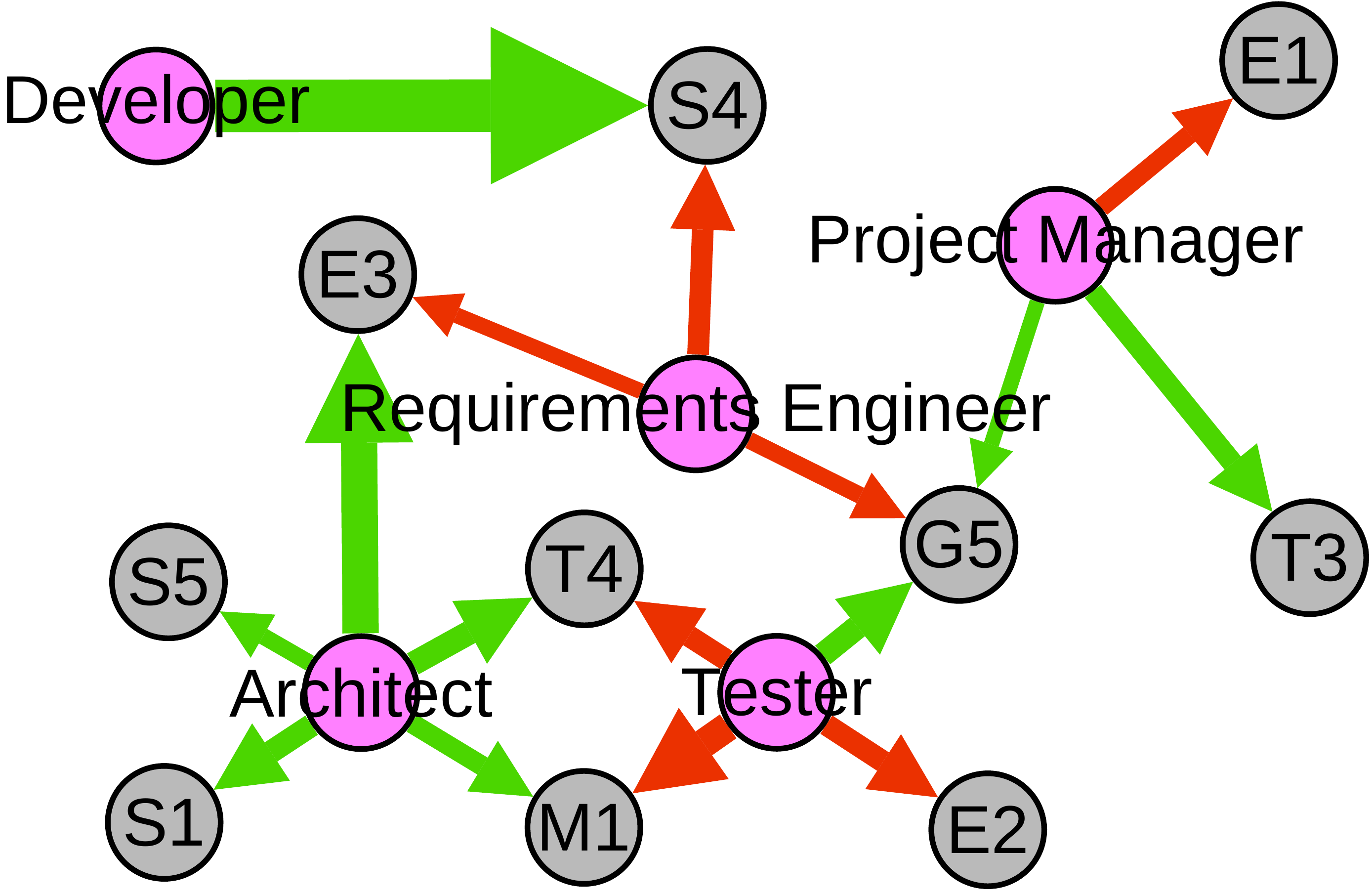}
	      \label{fig:contextfactor:role}	      
      }
      \subfloat[Agility]{
	      \includegraphics[width=0.3\textwidth]{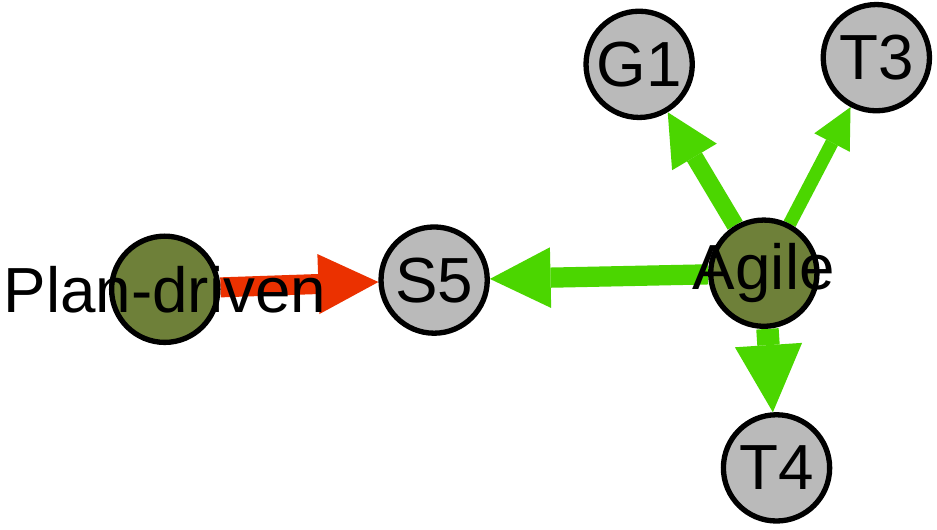}
	      \label{fig:contextfactor:agility}
      }
      \subfloat[Company size]{
	      \includegraphics[width=0.3\textwidth]{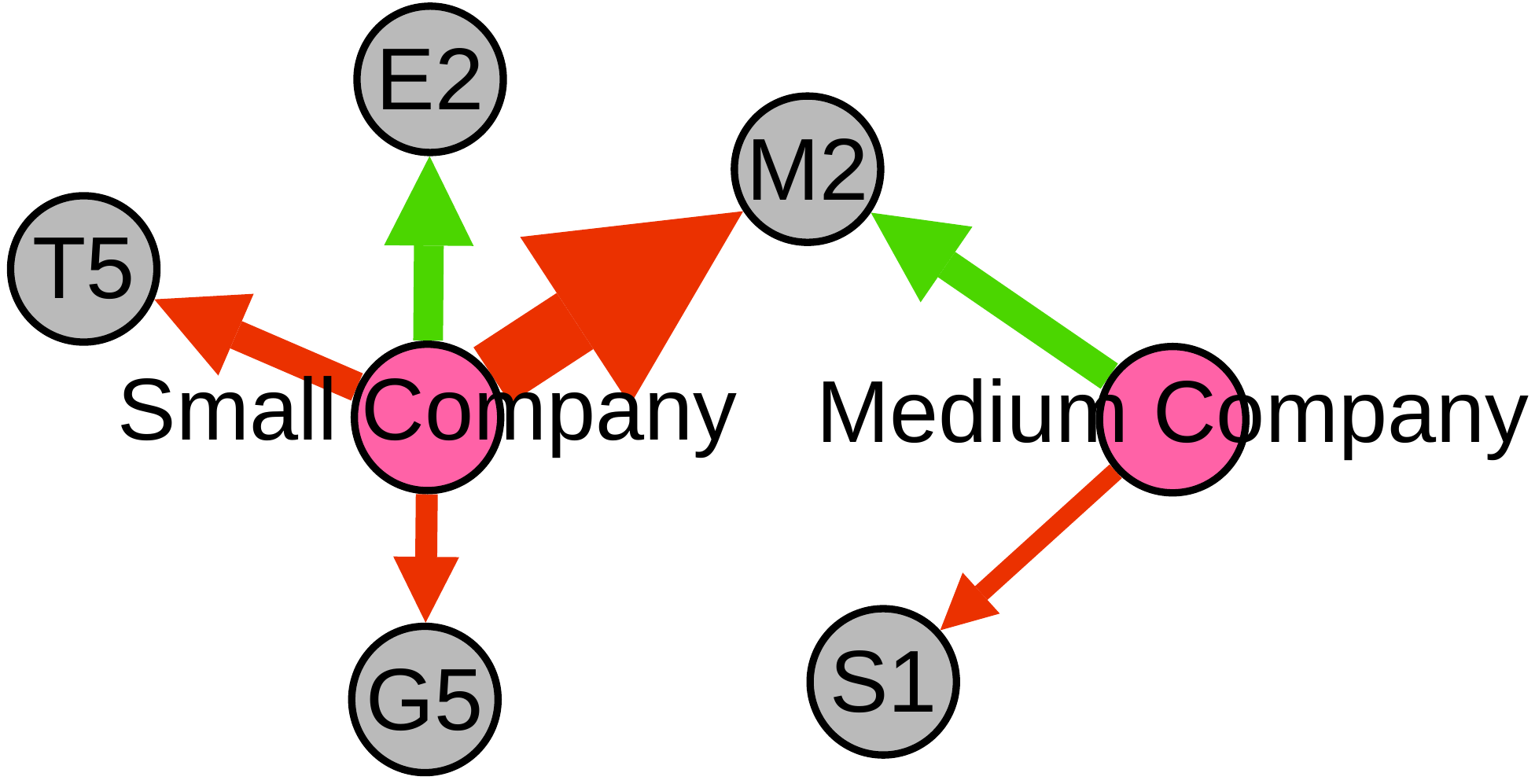}
	      \label{fig:contextfactor:companysize}
      }
      
      \subfloat[Experience]{
	      \includegraphics[width=0.15\textwidth]{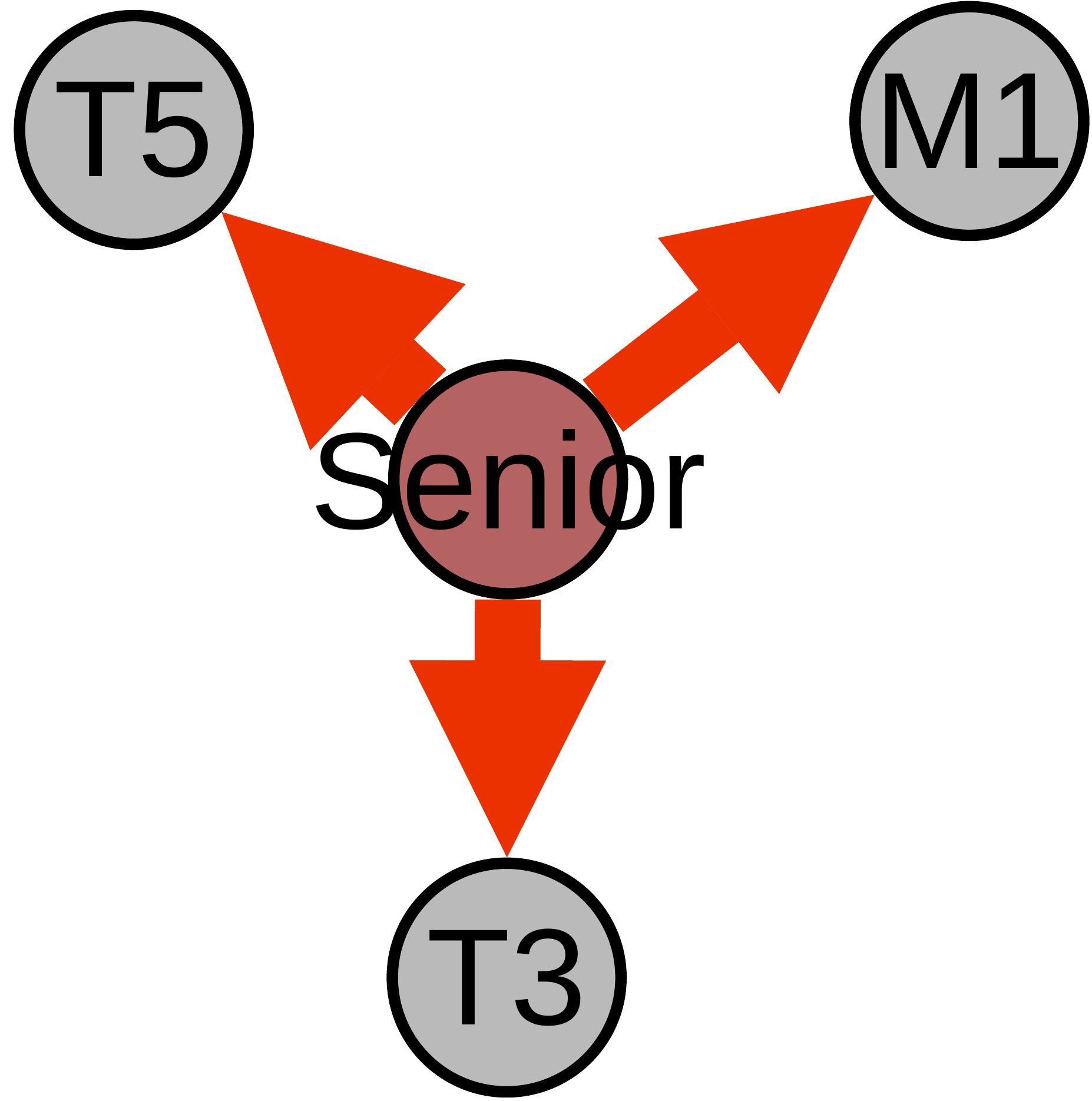}
	      \label{fig:contextfactor:experience}
      }
      \subfloat[QR documentation]{
	      \includegraphics[width=0.15\textwidth]{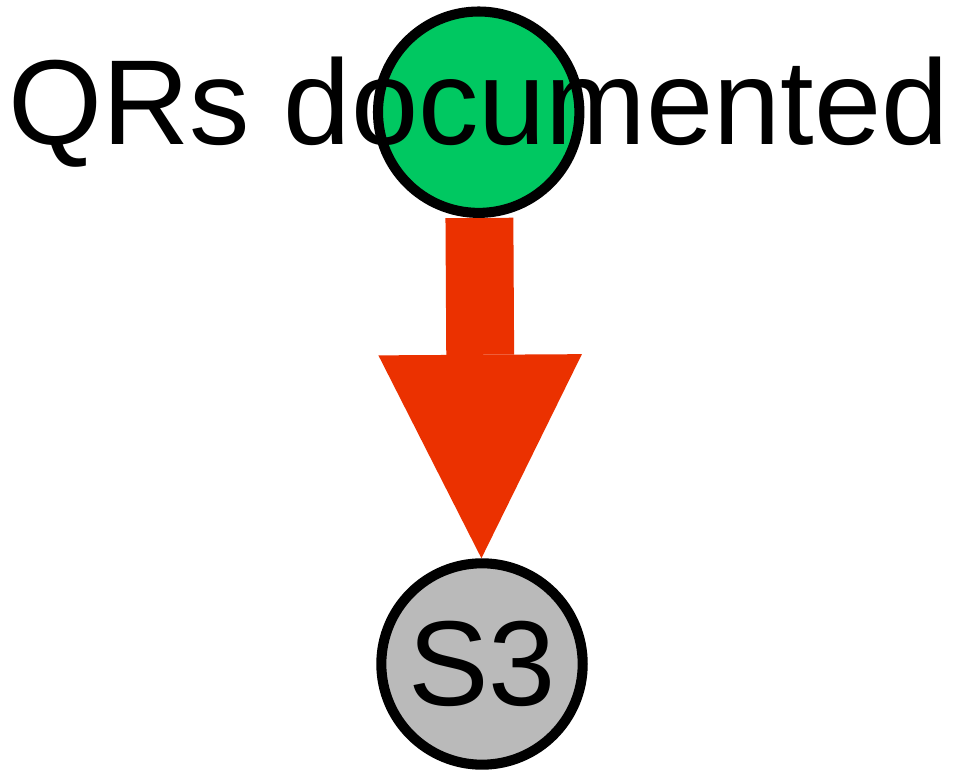}
	      \label{fig:contextfactor:qrdocumentation}
      }
      \subfloat[Company distribution]{
	      \includegraphics[width=0.3\textwidth]{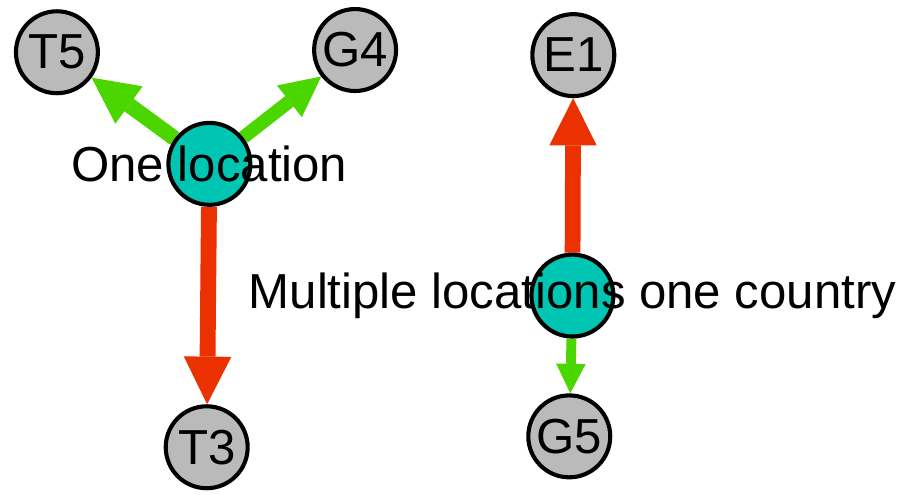}
	      \label{fig:contextfactor:distribution}
      }
      \subfloat[Domain]{
	      \includegraphics[width=0.4\textwidth]{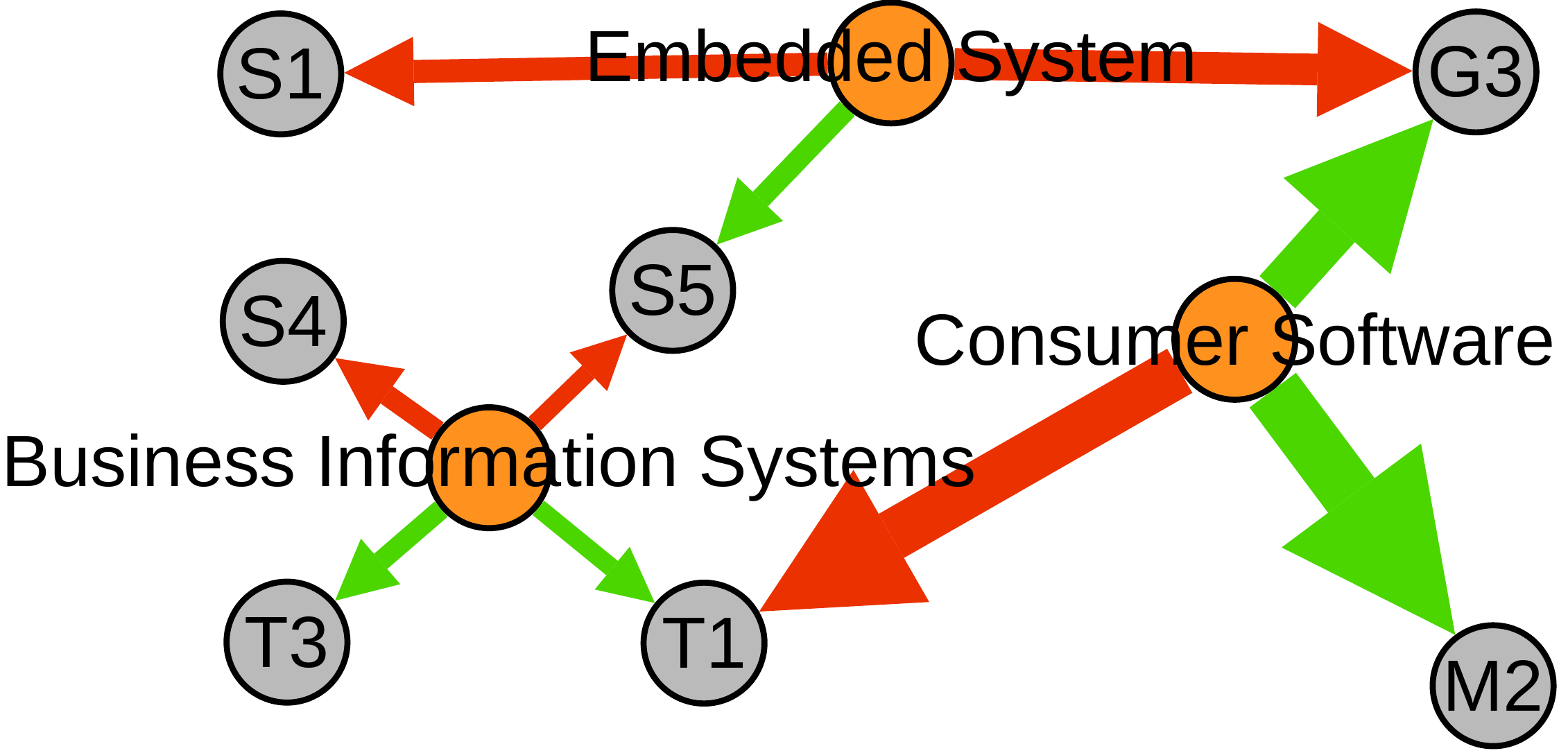}
	      \label{fig:contextfactor:domain}	      
      }
      
   \caption{Context factor analysis (1/2): The research statements are depicted as gray nodes and the context factors as colored nodes. Edges indicate a tendency to agree more (green) or to disagree more (red). The width of the arrow indicates the value of the regression coefficient.}

  \label{fig:contextfactoranalysis1}   
\end{figure}

\begin{figure}
   \centering
     \subfloat[Project Size]{
	      \includegraphics[width=0.35\textwidth]{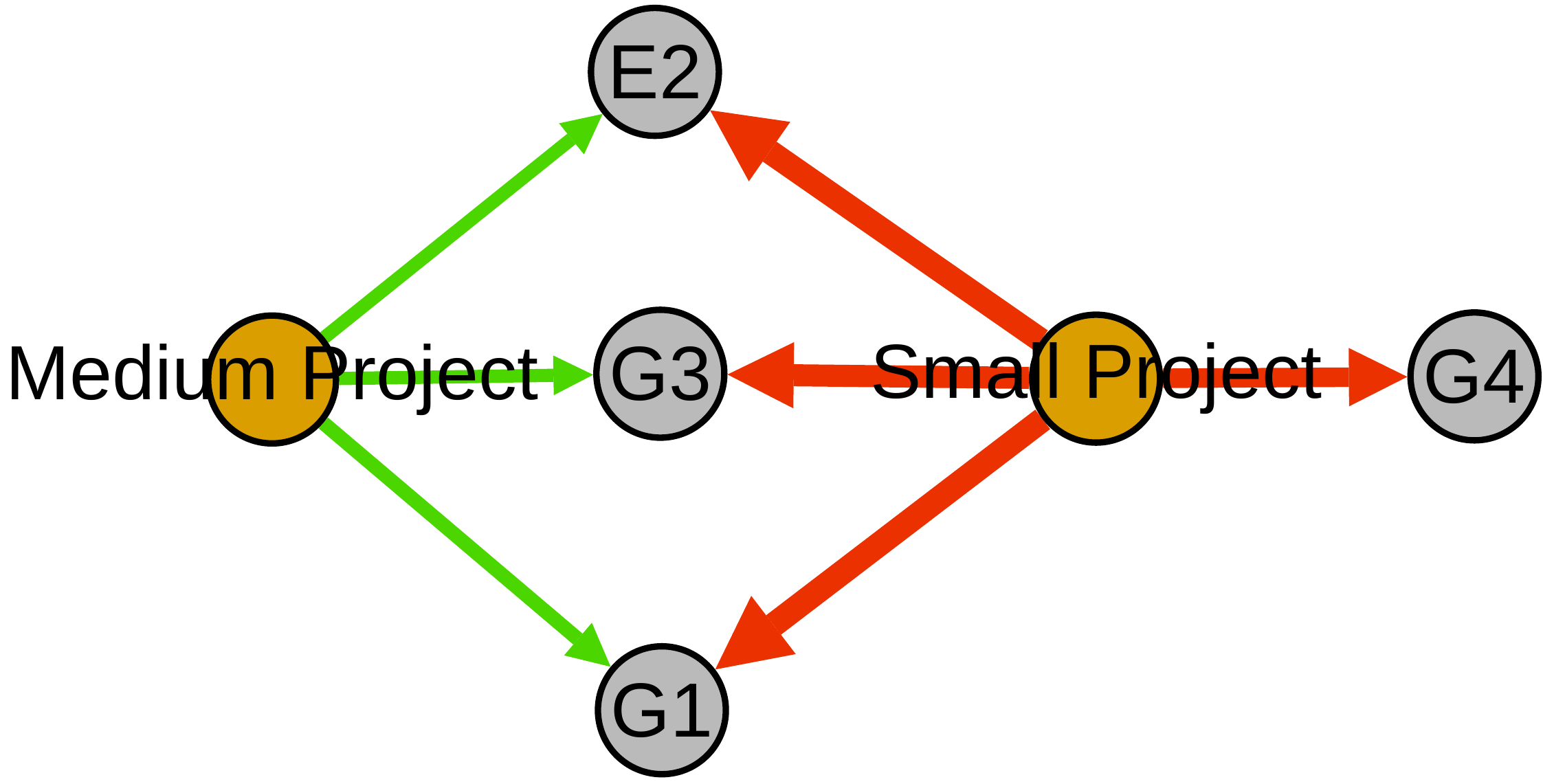}
	      \label{fig:contextfactor:projectsize}
      }
      \subfloat[SRS role]{
	      \includegraphics[width=0.25\textwidth]{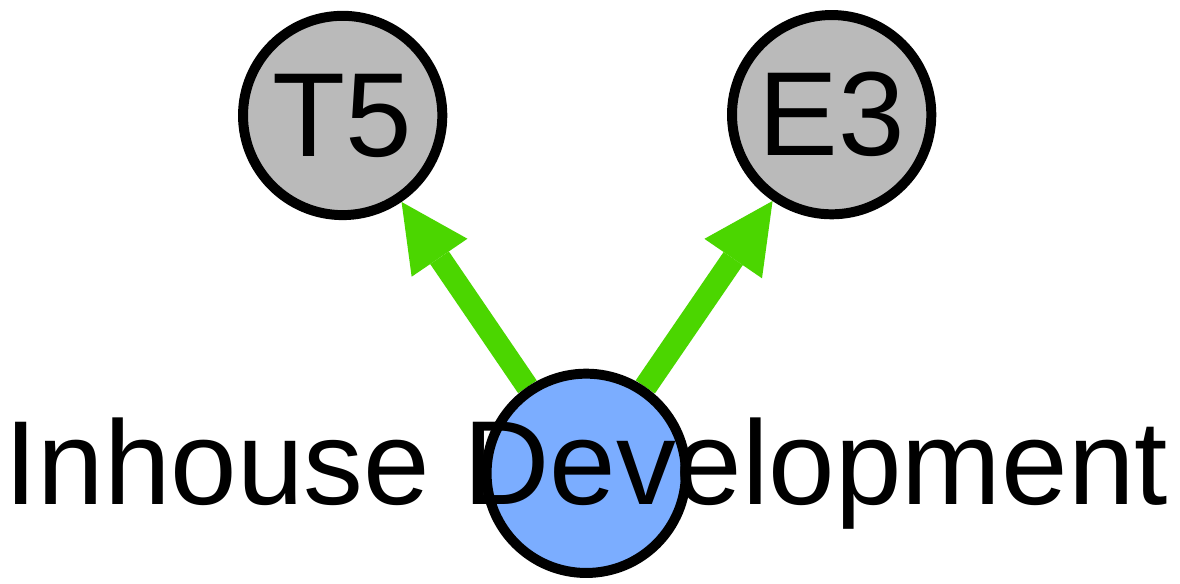}
	      \label{fig:contextfactor:SRSrole}
      }
      
      \subfloat[QR Importance]{
	      \includegraphics[width=0.8\textwidth]{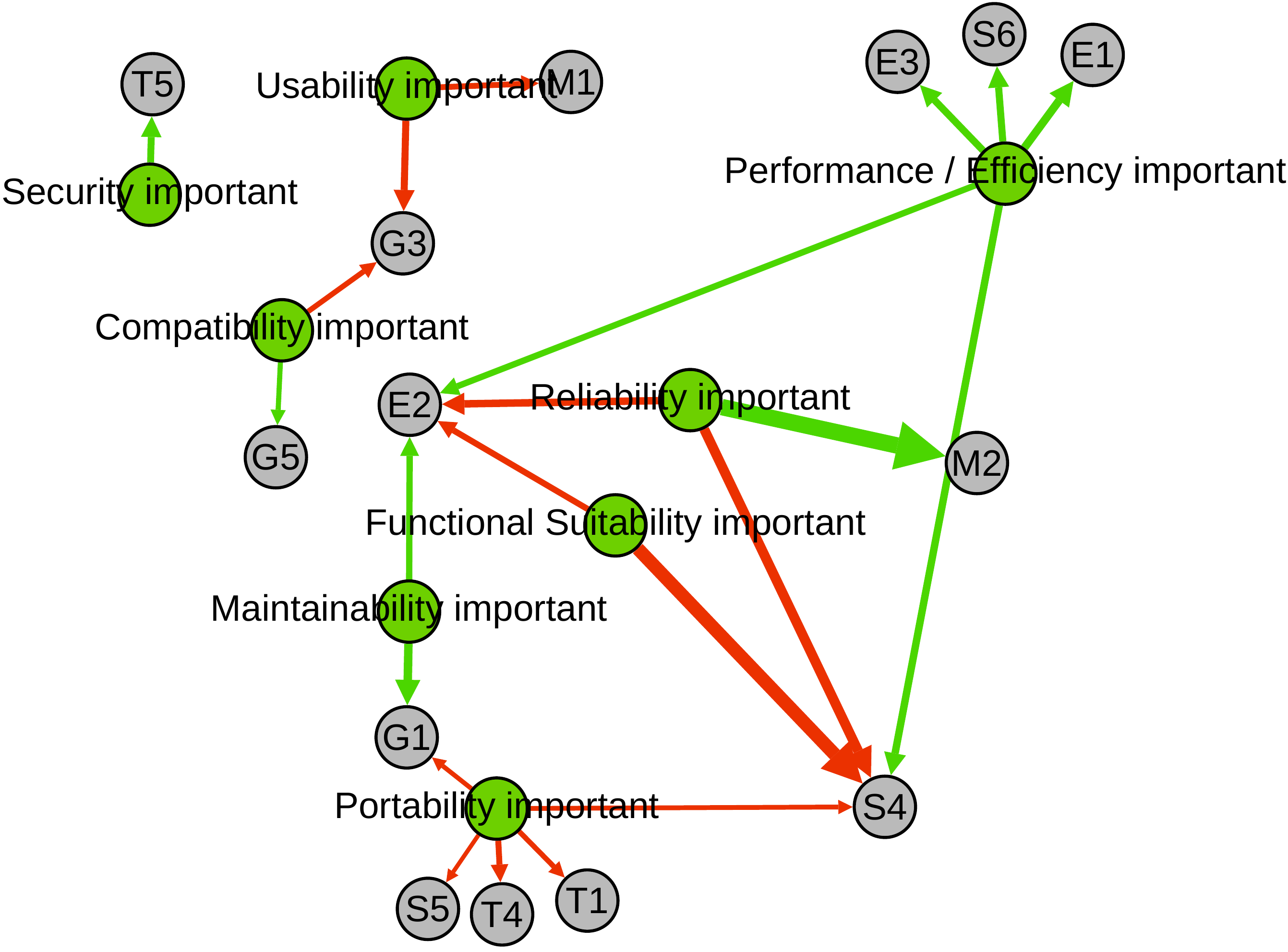}
	      \label{fig:contextfactor:qrimportance}
      }

      \subfloat[Sector]{
	      \includegraphics[width=0.7\textwidth]{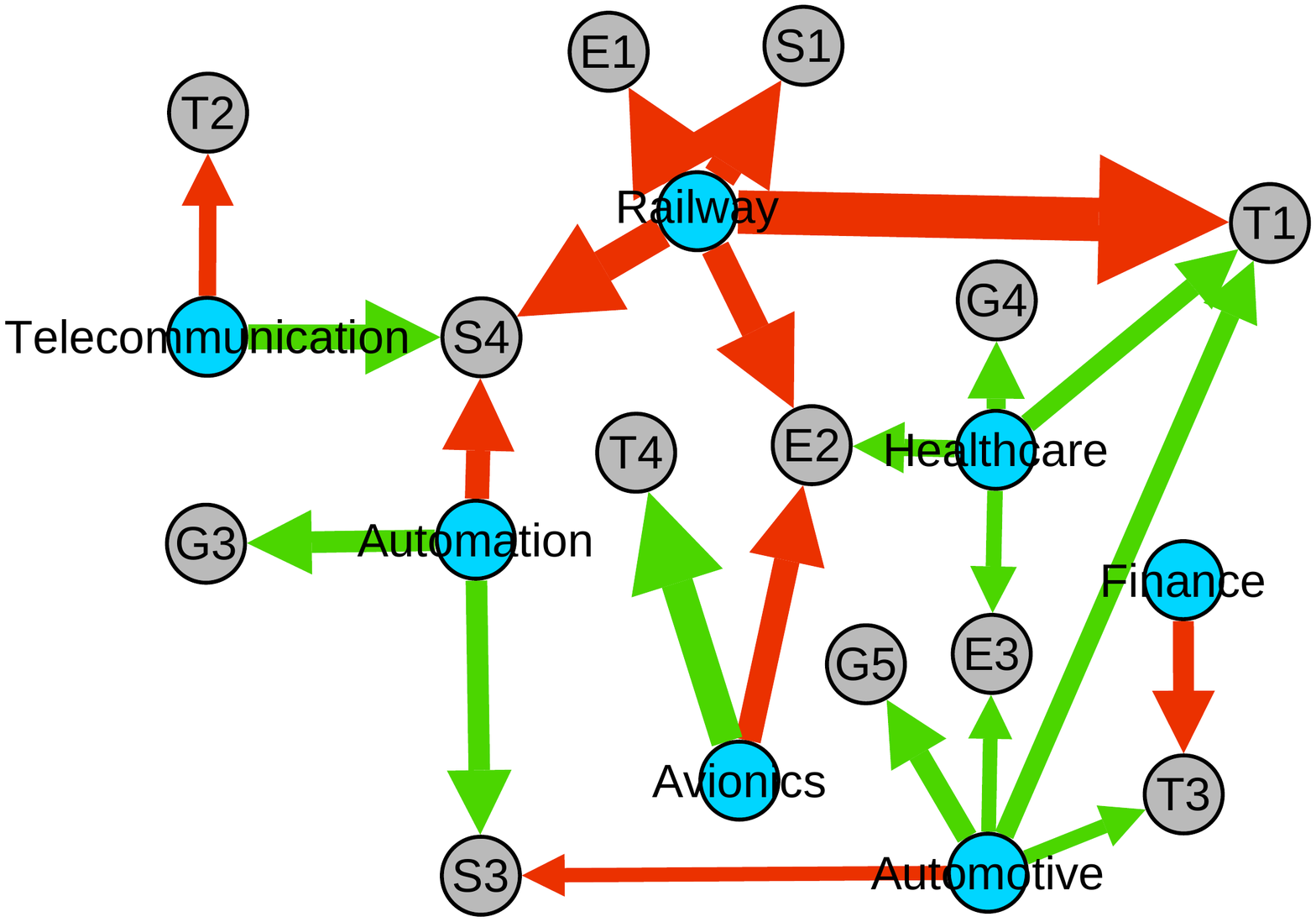}
	      \label{fig:contextfactor:sector}
      }
   \caption{Context factor analysis (2/2): The research statements are depicted as gray nodes and the context factors as colored nodes. Edges indicate a tendency to agree more (green) or to disagree more (red). The width of the arrow indicates the value of the regression coefficient.}
  \label{fig:contextfactoranalysis2}   
\end{figure}

To better understand the influence of context factors on agreement or disagreement, we also report the results of our regression model along the significant context factors. In particular, we create a graph that contains all research statements (G1-5, E1-3,S1-6, T1-5, M1-2) as nodes and all context factors that show at least one significant influence in our regression model as nodes. For each context factor that shows a correlation with agreement or disagreement with a research statement, we add an edge to the graph, colored in green (positive) and red (negative), respectively. Moreover, we weight the width of each edge with the value of the regression coefficient. Since the whole graph with all statements and context factors is rather large,
\Cref{fig:contextfactoranalysis1,fig:contextfactoranalysis2} show the subgraphs for each group of context factors.\footnote{The \emph{gephi} file of the whole graph is part of our additional material package~\cite{Vogelsang18}.} 

For example, in \Cref{fig:contextfactor:role}, the relationship between the role of a participant and the research statements is shown: Testers tend to disagree more with research statements M1, E2, T4 and tend to agree more with research statement G5.

\subsection{Summary and Categorization of Results}
\label{sec:res_summary}
\begin{figure}
\centering
  \includegraphics[width=\textwidth]{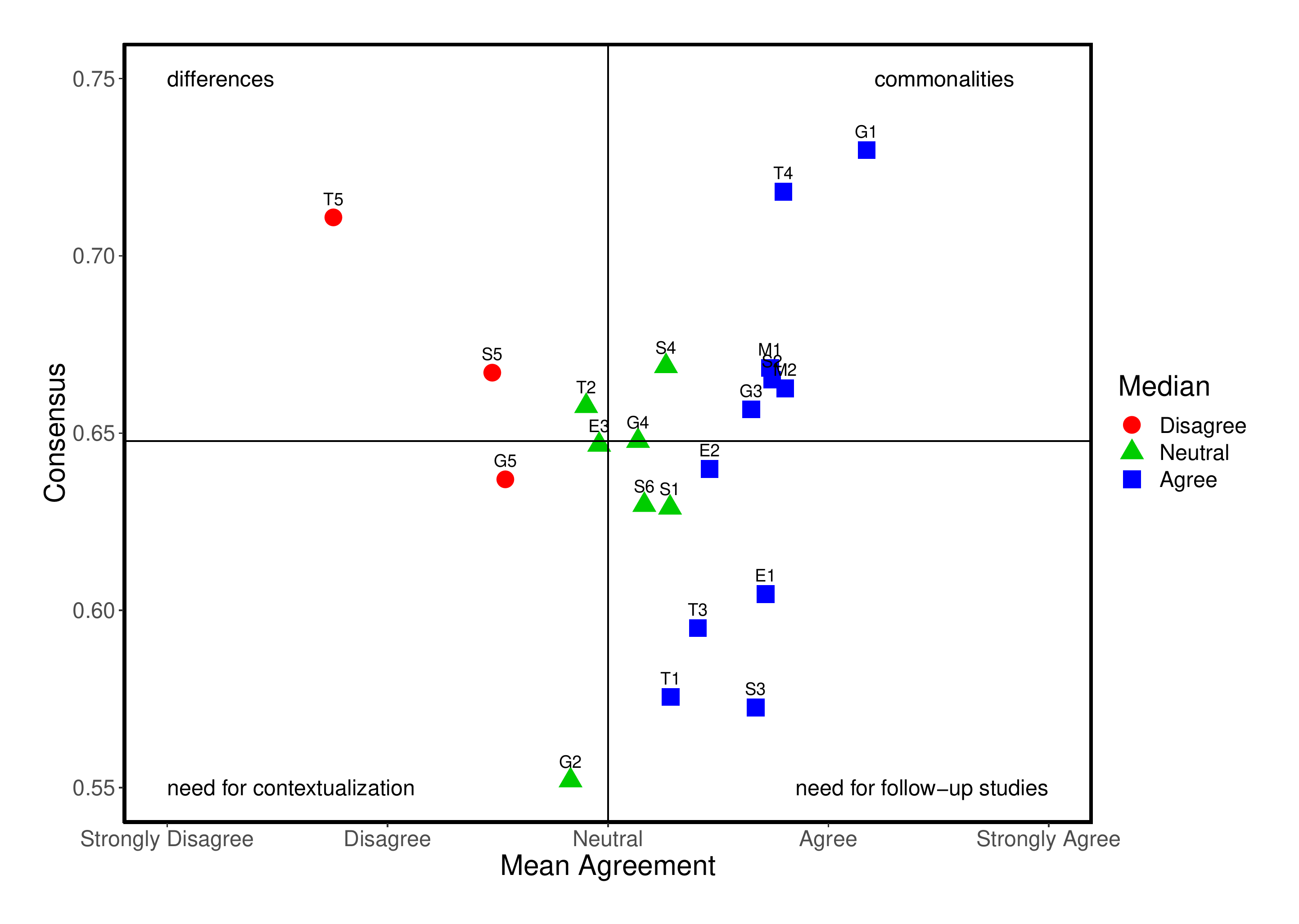}
   \caption{Classification of research statements into four areas based on consensus and mean agreement.}
  \label{fig:sdmean}
\end{figure}

In \Cref{fig:sdmean}, we have plotted the research statements with respect to their mean level of agreement\footnote{We are aware that using the mean as a measure for central tendency of Likert (i.e., ordinal) scales is something to be careful with. Therefore, we refrain from interpreting the mean value itself but use the value only to order the statements with respect to their level of agreement. In addition, we have labeled the statements also with respect to their median value.} and their consensus value, which is a measure of dispersion for answers on Likert scales~\cite{Tastle07}. 
We divide this plot into four areas: We consider statements with a high level of agreement and high level of consensus as \emph{Commonalities} (between academic statements and practitioners' perception). 
In contrast, we consider statements with a low level of agreement and high level of consensus as \emph{Differences} (between academic statements and practitioners' perception). 
Research statements with a low level of consensus are more interesting for research. If the general level of agreement is high but there is also low consensus in the answers, there is a need for follow-up studies to investigate why the statements may not be true in certain areas (i.e., why some specific respondents disagreed with the statements). If the general level of agreement is low and there is also a low level of consensus in the answers, there is a need for investigating the context of the original studies in more detail. It might be the case that the statement has been stated in a particular context that does not generalize to other contexts. 

In \Cref{tbl:categories}, we list the four categories with the belonging statements. In the table, we list only the statements with a non-neutral median agreement.

\begin{table}
   \caption{Summary of research statements and categories}
  \label{tbl:categories}
\centering
  \begin{tabularx}{\textwidth}{@{}X@{}}
  \toprule
  \textbf{Commonalities}\\
  G1: The application domain strongly influences the relevance of individual types of QRs \\
  T4: Testing QRs is time-consuming\\
  M1: QRs are often not sufficiently prioritized\\
  S2: The documentation of QRs is not always precise\\
  M2: Software architects do not use a specific tool for QR management\\
  G3: QRs are sometimes ignored\\
  \midrule
  \textbf{Differences}\\
  T5: Testing QRs is impossible\\
  S5: Functional requirements are often labeled as QRs\\
  \midrule
  \textbf{Need for follow-up studies}\\
  S3: QRs are often described in too vague terms\\
  T1: Only few types of QRs are validated at the end of the project\\
  T3: Most QR types are difficult to test properly\\
  E1: In requirements elicitation, the focus is on FRs, not on QRs\\
  E2: Many QRs remain undiscovered\\
  \midrule
  \textbf{Need for contextualization}\\
  G5: Only few QRs deal with architectural aspects\\
  \bottomrule
  \end{tabularx}
\end{table}

\subsubsection{Commonalities}
We identified six statements with a generally high level of agreement among the participating practitioners. 
Those statements include 
G1 (``The application domain strongly influences the relevance of individual types of QRs''), 
T4 (``Testing QRs is time-consuming''), 
M1 (``QRs are often not sufficiently prioritized''), 
S2 (``The documentation of QRs is not always precise''), 
M2 (``Software architects do not use a specific tool for QR management''), 
and G3 (``Many QRs describe functional aspects of a system'').
These statements are not only commonly agreed between academics and practitioners with a high level of consensus, but they seem also to be commonly agreed within the academic research community itself as we found only one publication that argues against one of these statements (M1). 
For all other statements, we only found publications supporting the statements (see \Cref{tbl:researchStatementsList}).

\subsubsection{Differences}
We identified two statements with a low level of agreement and a high level of consensus, i.e., these statements are rejected by practitioners in general. This area includes T5 (``Testing QRs is impossible'') and S5 (``Functional requirements are often labeled as QRs'').
The general disagreement with statement S5 is interesting since we found three independent publications~\cite{Eckhardt16,Ernst10,Rahimi14}  where this statement was issued (see \Cref{tbl:researchStatementsList}).

One explanation we have for the strong disagreement with statement T5 is, not surprisingly, its universal nature. It is reasonable to assume that our respondents have encountered at least one situation where this statement does not hold, especially when considering quality requirements for which testing procedures have, in fact, been adopted in one form or the other; for instance, because of their functional nature that allows for the direct adoption of existing testing procedures (e.g., performance-related requirements), but also because other QR classes are often in scope of dedicated engineering activities (e.g., usability-related requirements being in scope of, for example, prototypes and mock-ups).

\subsubsection{Need for Follow-up Studies}
Statements for which we identified the need to conduct follow-up studies are those with a high-level of agreement while, at the same time, showing a low level of consensus among the responding practitioners. 
For example, the statement E1 (``In requirements elicitation, the focus is on FRs, not on NFRs'') has a high level of agreement in general but a low level of consensus. 
In terms of related publications, this statement has a strong standing with six papers supporting the statement and none opposing it (see \Cref{tbl:researchStatementsList}). 
However, respondents from the railway sector, for instance, have strongly objected this statement (see \Cref{tbl:researchStatements}). Therefore, it may be interesting to investigate this domain in more detail to find out why the statement seems not to be true in that domain.
We see similar trends for statements S3, T3, and E2. For all of these statements, we found only publications supporting the statements, however, \Cref{tbl:researchStatementsList} shows certain groups of practitioners who disagree with the statements (e.g., automotive sector for S3, senior developers for T3, or railway sector for E1 and E2).

\subsubsection{Need for Contextualization}
We identified only statement G5 (``Only few NFRs deal with architectural aspects'') in this particular area. It is the only statement with a low level of agreement between practitioners in general but also with a low level of consensus in the answers. That could mean that the statement was stated in a specific context where it may be true but in many other contexts, the statement is perceived as not true. This is in tune with the balanced number of supporting and opposing publications related to this statement (see \Cref{tbl:researchStatementsList}).

\section{Discussion}
\label{sec:discussion}
In the following, we discuss the results that we found particularly interesting and contrast them with the original studies where they were stated.

Please note that we do not discuss the results of every research statement in detail as this would be purely speculative. Hence, the presented discussion can only serve as starting point for further studies examining the discussed aspects because our study does not provide any data on the specific reasons why practitioners agree or disagree with a research statement.

\subsection{In Elicitation, Priority is on FRs, not on QRs}
Most of our respondents confirmed that in requirements elicitation, the focus is on FRs, not on QRs (E1). Respondents from the railway domain and from large, distributed projects are an exception. They tend to disagree more with this statement. Nevertheless, most respondents confirmed that QRs are at least documented in the end. They disagreed with the statement that QRs are not documented at all (S1). This is also in line with observations from another survey, where 88\% of respondents answered that they document QRs~\cite{Eckhardt17NFRvsFR}. As a result of this lower priority of QRs in the elicitation, our respondents stated that many QRs remain undiscovered (E2) and only few types of QRs are validated at the end of a project (T1). Interestingly, respondents from the railway sector who stated that the elicitation focuses not necessarily on FRs only also disagreed with the negative consequences of undiscovered and not validated QRs. Further studies are neccessary to disover a possible causal relation between these effects. In general, practitioners seem to focus on FRs in requirements elicitation and handle QRs with lower priority.

\subsection{Different Roles have Diverging Opinions about QRs}
% G5 (Only few arch. aspects): Manager, Tester pos, RE neg
% E3 (elicitatio by arch): Architect pos, RE neg
% S4 (doc becomes desynch): Developer pos, RE neg
% T4 (testing is time consuming): Architect pos, Tester neg, 

As shown in \Cref{fig:contextfactor:role}, respondents with different roles have different opinions (regarding agreement\slash disagreement) on the research statements. Specifically, requirements engineers and architects seem to disagree on the responsibility of QR elicitation. While architects overly agreed that QR elicitation is mainly done by them (E3), requirements engineers tend to disagree with this statement. A similar divergence is between requirements engineers and developers when it comes to documentation of QRs (S4). It seems that developers perceive QR documentation more often as not in sync with the current state of the system, while requirements engineers do not perceive this threat as that strong.

Testing QRs is perceived as time-consuming especially by architects. Testers, on the other hand, overly disagreed with the statement T4 (\emph{Testing QRs is time-consuming}). This seems an interesting point for further research on specific tests that architects have in mind.

The statement that only few QRs deal with architectural aspects (G5) is interesting since the architects themselves do not have a strong opinion for or against that statement while all other roles have. Testers and managers agree more with the statement while requirements engineers disagree more.

\subsection{Are QRs really Non-Functional?}
Research statements G2 and S5 belong to the statements with which the respondents disagreed most (39\% and 59\% disagreement). Both statements address the confusion about the classification of requirements. G2 states that many QRs describe functional aspects of a system and S5 states that functional requirements are often labeled as QRs, which sounds counterintuitive. One explanation to us is that this could be the reason why respondents had a tendency to disagree with these. The statements originate from a paper where the authors investigated requirements specifications from industry~\cite{Eckhardt16}. In the documents they analyzed, they found a considerable number of requirements that were labeled as QR but that actually described functionality. The disagreement with the statements shows that the results of that study are controversial. Therefore, we need more studies that investigate the effects of blurry classification rules.

\subsection{Testing QRs is hard but not impossible}
In general, our respondents were not as pessimistic about the possibility to test QRs as the original study suggests. In fact, 86\% disagreed with the respective statement T5 (\emph{Testing QRs is impossible}). This shows that our respondents at least seem to have some ideas how to test QRs. Especially experienced respondents overly disagreed with this statement. Similarly, respondents from large companies strongly disagreed. It would be interesting to see what impact the company size has on testing procedures. Respondents that tend to agree more with the impossibility of testing QRs worked in non-distributed projects or considered security as an important quality attribute for their systems. This may indicate to that we are still lacking good testing mechanisms for security. 

On the other hand, our respondents support the statement that testing QRs is difficult (T3) and time-consuming (T4). 55\% agreed or even strongly agreed that testing QRs is difficult. Only 8\% disagreed with the statement that testing QRs is time-consuming, while 18\% even strongly agreed with it. Especially in the avionics and automotive industry, testing seems to be a big issue since respondents of these industries overly agreed with T3 and T4. 
Also, respondents who stated that they work with an agile process paradigm agreed significantly stronger with the difficulty and necessary effort to test QRs.
The only group of respondents that tend to disagree more with the statement that testing is time-consuming are the testers themselves. It seems that testers have a more optimistic perception towards testing QRs.

In the original publication of statements T4 and T5, the authors stated a refined version of the statements. They said ``[\ldots] when expressed in non-measurable terms [QR] testing is time-consuming or even impossible''~\cite{borg2003bad}. In our survey, this premise about a measurable specification of QRs does not play a specific role for a stronger agreement. The context factors that correlate with a strong agreement to statements T4 and T5 do not correlate with the research statements that relate to a precise documentation of QRs (e.g., S2, S3).

\subsection{Different Domains, Different Problems}
For a number of research statements, the industry sector of the respondents is strongly correlated with the participant's agreement to the statement. This is further corroborated by research statement G1 (\emph{The application domain strongly influences the relevance of individual types of QRs.}) showing the strongest agreement among the respondents. This means that the perceived importance, the handling, and the problems associated with QRs strongly depend on the industry sector. For instance, while respondents from the healthcare and automotive sector strongly agreed with a number of research statements (see \Cref{fig:contextfactor:domain}), the respondents from the railway sector strongly disagreed with a number of research statements: research statement T1 (\emph{Only few types of QRs are validated at the end of the project.}) seems to be strongly accepted in the healthcare and automotive domain but not in the railway domain. Similarly, research statement E2 (\emph{Many QRs remain undiscovered}) was well received by respondents from healthcare but opposed by respondents from avionics or railway. This could indicate to the different relevance of QRs in the respective domains. Railway, healthcare, and automotive may provide interesting domains for conducting further case studies on the handling of QRs because respondents from these areas had strong positive or negative feelings about the research statements.

\subsection{QRs and Architecture: A Love-Hate Relationship}
% In disem Block gehe ich auf folgende Statements ein
% G4: common terminology, 
% G5: Only few QRs deal with architectural aspects, 
% M2: no specific tool,
% E3: Elicitation by architects
Many publications stress the importance of QRs for architectural decisions. The participant's answers support this relation. The statement that only few QRs deal with architectural aspects (G5) was rejected by the majority of respondents (56\%), 
especially by respondents from small companies and by requirements engineers. On the other hand, there were also some groups of respondents that significantly agreed more with this statement, namely respondents from the automotive industry, testers, and managers. It would be interesting to investigate what these groups see in QRs besides architectural aspects. In the original publication of that statement~\cite{Eckhardt16}, the authors reported that they identified a large number of requirements that they found in QR sections of industrial requirements documents as functional requirements.

Despite this important relation between QRs and architecture, the respondents mostly agreed with a lack of common terminology between architects regarding QRs (G4) and also with the statement that there is no specific tool support for managing QRs (M2). Yet, these issues do not seem to be that prevalent in small projects or companies as respondents from these groups disagreed more with statements G4 and M2.

\subsection{The Perceived Importance of Quality Attributes Shapes the Opinion}
% Performance:
% S4 agree S6. pos, E1: agree, E2 agree, e3 diagreee
% 
% Portability:
% S5 disagree, G1 disagree, T4 disagree, s4 disagree, t1 disagree
%
% Reliability:
% E2 disagree, s4 disagree, m2 agree

In the results, the agreement tendency also correlates with the perceived importance of QR types for the systems that the respondents develop.

Respondents who considered performance\slash efficiency as an important quality in their systems also had a distinct opinion about elicitation and specification of QRs. For example, these respondents overly agreed with the statement that in elicitation, the focus is on FRs and not on QRs (E1). On the other hand, they also tend to agree more with the statement that many QRs remain undiscovered (E2). Additionally, respondents disagreed that QRs are mainly elicited by architects (E3). The authors of another study on a similar topic mention that companies have specialized teams focusing on performance or reliability~\cite{Eckhardt17NFRvsFR}. This could indicate that QR elicitation is mainly done by these specialized teams. A downside of that could be that the documentation of QRs becomes desynchronized to which our respondents with focus on performance also agreed overly (S4).

The opinion of respondents who considered portability as an important quality in their systems deviated several times from the average opinion of all respondents. They overly disagreed with the influence of the application domain (G1). Interestingly, people who considered maintainability as an important quality attribute overly agreed with the influence of the application domain. The study of Jung~et~al.~\cite{Jung04} provide indication that portability and maintainability actually measure the same intrinsic concept. However, this study was only performed in the context of one specific system. Our study indicates that there may still be a difference between the two characteristics in some application domains. 
Regarding specification of QRs, respondents who considered portability as an important quality overly disagreed with the statement that functional requirements are often labeled as QRs (S5) and with the possible desynchronization of QR documentation (S4). In testing, they disagreed that testing is time-consuming and that only few types are validated at the end of a project. This could indicate that testing of systems with a high demand of portability is actually in a good shape.

\subsection{Experience Fosters Optimism}
% Seniors haben besondere Meinung zu den folgenden Statements:
% T5: Testing is impossible
% M1: NFRs are not sufficiently prioritized
% T3: Difficult to test

When focusing on the group of respondents that are already experienced in their field (more than 3 years of experience), we see that these disagree more with some of the negative statements about QRs. For example, seniors are not so pessimistic about the possibility of testing QRs (T5), the difficulty of testing QRs (T3), and the insufficient prioritization of QRs (M1). These results may indicate that testing and prioritizing QRs is hard and need practice to be done properly. On the other hand, this may also indicate that we currently lack sufficient methods and tools for these activities, which seniors may compensate with experience and best practices. 

A second interpretation of these results is that the different perceptions are due to a changing perception of related risks. Novices may still strive for perfection (e.g., 100\% test coverage) and therefore perceive testing and prioritizing QRs as overly hard. Seniors, on the other hand, may have experienced that small deviations from a perfect solution can be compensated and, thus, perceive some of the challenges related to QRs as less dramatic.

\subsection{Scale Changes the View on QRs}
% The smaller project, the more disagree:
% E2: keine starke Tendenz
% G3, keine starke Tendenz
% G1: keine starke Tendenz
% 
% M2: larger companies -> more agree
% mit Tendenz

% E2 ist interessant weil small projects disagree aber small companies agree

Working in small projects and small companies is different from working in large and distributed projects. Our results show that this difference also correlates with the perception of some QR statements. As for the project size, we found three statements (G1, G3, and E2) with which respondents in small projects overly disagreed while respondents from medium and large projects overly agreed. In small projects, the influence of the application domain on the relevance of QR types is perceived less strong compared to larger projects (G1). Additionally, it seems that some problems related to QRs get more serious in larger projects. Respondents from medium and large projects overly agreed that many QRs remain undiscovered (E2) and that QRs are sometimes ignored (G3), while respondents from small projects overly disagreed with these statements.

Interestingly, the effects of differences in project size are not in tune with those of differences in company size. For company size, only research statement M2 shows a tendency that respondents of small companies disagreed stronger with the statement that architects do not use a specific tool for QR management while respondents from medium and large companies overly agreed with this statement. For statement E2 (\emph{Many QRs remain undiscovered}), we even got a counterintuitive result since respondents from small projects overly disagreed with the statement, while respondents from small companies overly agreed with it. This means that undiscovered QRs are not so much of a problem in small projects, however, they are in small companies.

\section{Threats to Validity}
Our results and conclusions are subject to a number of threats that we discuss in the following:

\subsection{Identification of Research Statements}
The identification of the research statements as described in \Cref{sec:study_design:statements} has not been done as part of a systematic literature study. One of the authors browsed through the papers of major recent requirements engineering and software engineering venues to identify papers on quality requirements first and then extract statements from them. Since we validated and discussed the extracted statements in the team of authors, we are confident that the analyzed statements fit the study purpose; however, we do not claim that the list of research statements is complete or representative for all views in the research community. In fact, we have to assume that there are other research statements on QRs that we have not addressed in our study. Therefore, we refrain from making any statements about the completeness of research statements on QRs in general. Nevertheless, we argue that this does not invalidate the results we gathered from the research statements in scope. It is reasonable to doubt that respondents would have answered differently if there were more research statements.

\subsection{Participant Selection} One limitation in the study is the missing lack of control over the respondents given that we distributed the survey invitation over various networks. Apart from the ultimately unknown response rate, this means that we cannot control how representative the responses are. Despite the applied validity procedures described in \Cref{sec:study_design:v-procedures}, we cannot guarantee that all the views taken really result from practitioners.

When looking at the distribution of context factors, we see that some context factors are only represented by a low number of respondents (e.g., railway domain, junior engineers, or consumer software). This does not invalidate the significance verdicts of the Wald test because smaller sample sizes may lead to larger confidence intervals and, thus, increase the likelihood of non-significant results. However, small sample sizes, on the other hand, may decrease the statistical power of our results. Lower statistical power indicates a lower probability that an effect is actually true when the Wald test says it is significant.
In other words, the small sample sizes for some context factors do not threaten the validity of the observed significances but increase the chance that these significant results do not indicate a true effect because the few respondents with that specific context factor may not be representative for all subjects in that group.

\subsection{Survey Research} Further threats to the validity result from the nature of survey research. 
We cannot control on which basis the respondents provide their answers and the respondents might be biased. 
Secondly, there is the possibility that respondents have misinterpreted some of the questions or even the concept of QRs. Some of the research statements that we found in the literature are formulated very fuzzy (e.g., only few QRs deal with architectural aspects; How many are few? What are aspects?). Therefore, respondents may have interpreted the statements differently. Thus, respondents could select ``Don't know'' if they could not understand a statement. However, this option was selected in a few cases only (see also \Cref{fig:answerDistribution}). Our interpretation is that the respondents had an opinion on most of the statements in their exact wording.

\subsection{Data collection} In the questionnaire, we characterized some context factors by specific boundaries (e.g., large companies are those with more than 2,000 employees, seniors are those respondents with more than 3 years of experience). We set these boundaries based on what we believe is reasonable and what has been also used in other studies. However, the results may change if we would have set the boundaries differently.

\subsection{Statistical Model} The regression model applied here is well tailored to the analysis of the research statements measured on a Likert scale, as it exploits the ordinal structure of the answers. An attractive property is the simple interpretation of its parameters, which results from the proportional odds assumption. However, one should be aware that this implies a strong assumption on the underlying data generating process; that is, it assumes that the interpretation of parameters does not depend on the category, i.e., the level of agreement (see \Cref{sec:regression_coefficient}). On the other hand, an extended version of the model that we used in which all (or part of) the parameters are category-specific makes the parameters hard to interpret and in the presence of only 103 respondents yields unreliable estimates. 

\subsection{Limitations in the Generalization} Related to the threats described above and as it is often the case with survey-based research, the ability to generalize is difficult due to the size of the sample and due to the diversity of the answers given. The lack of related studies with explicit context factors similar to those captured in our survey further renders generalization by analogy very difficult, if not impossible. However, while generalization of certain assessments of statements (i.e., the opinions and experiences made by practitioners) were not part of our overall objectives, we still argue that we can generalize from our results at least to a few selected context factors where our respondents commonly agreed in their assessments. More precisely, we are confident in that the assessments of the research statements for which we have a high level of agreement and a high level of consensus among the respondents (see the upper right corner in \Cref{fig:sdmean}) could be generalized, with caution, to those context factors that seemed to play a role. One such example is statement G1. On the other hand, those statements with a rather low consensus indicate that we need, at least, more detailed investigations on the context factors to have a more differentiated view that would allow for generalizations (e.g., G2).

\section{Conclusions}

In our study, we aim at assessing the perception of research statements about quality requirements in the view of practitioners. We want to identify potential research gaps for specific contexts, potential differences that raise the need for further investigations, or statements that are only true in certain contexts.
To this end, we identified 21 exemplary research statements as reported in academic literature and surveyed practitioners about their general agreement with the statements. We were not only interested in whether practitioners agree or disagree with the statements but also whether context factors have an influence on the perception. This shall allow us to pinpoint to research statements where additional research might be useful. 

Our results are based on an online survey yielding 103 responses from practitioners having a broad spectrum of different backgrounds. We analyzed the responses by means of a statistical regression model that calculates the factor by which the probability to agree or disagree more changes if a context factor changes. Our results show that a majority of the statements is also well respected by practitioners; however, not all of them. When examining the different groups and backgrounds of respondents, we noticed interesting deviations of perceptions within different groups that may lead to new research questions. Especially the perceptions of respondents with different roles may explain why communication and clarification problems about QRs may occur in practice. Most respondents perceive testing QRs as time-consuming and difficult, especially in agile contexts and in the automotive and avionics sector. Testers themselves have a more positive view on testing QRs. They overly disagreed with the more pessimistic statements about problems related to QRs. We additionally conclude from our results that it is not reasonable to speak about QRs in general because different types of QRs have very different characteristics and related challenges. The importance of specific types of QRs strongly influenced their perception on how to elicit, document, test, and mange QRs.

Overall, however, we need to differentiate in our findings between statements to which practitioners commonly agreed with a high level of consensus and statements where the consensus is lower (see \Cref{fig:sdmean}). While the first category of statements allows for a certain generalization of our findings, the latter is especially interesting for additional research as it highlights the need for better contextualisation. As future work, we therefore motivate and plan ourselves for further investigations in this direction. In particular, we argue for the need for further replications considering more context factors as well as triangulations in the inquiry methods going beyond survey research. Especially case study research is of particular interest to further explore the perceptions of practitioners on the research statements with respect to the particularities of their context.

Finally, one hope we associate with our work is also that other researchers are encouraged to describe more explicitly the context in which they conduct their empirical studies, and that they discuss the conditions and possible limitations of their statements. Otherwise, their conclusions might just contribute further to the existing leprechauns that still dominate requirements engineering research.

\section*{Acknowledgements}
We thank the respondents of our survey for sharing their opinion with us. We furthermore thank Wolfgang B{\"o}hm, Kevin Schlieper and Tobias M{\"u}hlbauer for piloting the study and Jan S{\"u}rmeli for feedback on earlier versions of this manuscript. 

\bibliographystyle{elsarticle-num}
\bibliography{bib}

\begin{thebibliography}{10}
\bibitem{DC06}
D.~Damian, J.~Chisan, An empirical study of the complex relationships between
  requirements engineering processes and other processes that lead to payoffs
  in productivity, quality, and risk management, IEEE Transactions on Software
  Engineering 32~(7).
\bibitem{Jansen10}
H.~Jansen, The logic of qualitative survey research and its position in the
  field of social research methods, Forum Qualitative Sozialforschung / Forum:
  Qualitative Social Research 11~(2).
\bibitem{Wald43}
A.~Wald, Tests of statistical hypotheses concerning several parameters when the
  number of observations is large, Transactions of the American Mathematical
  society 54~(3).
\end{thebibliography}

\end{document}